\documentclass[accepted]{uai2023}
\usepackage[american]{babel}

\usepackage[utf8]{inputenc}
\usepackage{csquotes}

\usepackage{amsmath}
\allowdisplaybreaks
\usepackage{mathtools}
\usepackage{bm}
\usepackage{amssymb}
\usepackage{listings}
\usepackage{soul}
\usepackage{natbib} 
\usepackage{hyperref}
\usepackage{algorithm}
\usepackage{algpseudocode}
\usepackage{subfig}
\usepackage{booktabs}
\usepackage{tabularx}

\newcommand{\vect}[1]{\mathbf{#1}}
\newcommand{\px}[1]{\cfrac{\partial #1}{\partial x}}
\newcommand{\py}[1]{\cfrac{\partial #1}{\partial y}}

\newcommand{\dt}[1]{\cfrac{\mathrm{d} #1}{\mathrm{d} t}}
\newcommand{\ds}[1]{\cfrac{\mathrm{d} #1}{\mathrm{d} s}}
\newcommand{\dnt}[2]{\cfrac{\mathrm{d}^{#1} #2}{\mathrm{d} t^{#1}}}
\newcommand{\Err}{\eta}
\newcommand{\Bound}{\mathcal{B}}
\newcommand{\Loss}{\mathrm{Loss}}
\newcommand{\Net}{\mathrm{Net}}
\renewcommand{\L}{\mathcal{L}}
\newcommand{\I}{\mathcal{I}}

\renewcommand{\Re}[1]{\mathcal{R}e\left(#1\right)}
\newcommand{\abs}{|\cdot|}

\title{Residual-Based Error Bound for Physics-Informed Neural Networks}
\author[1]{\href{mailto:<shuheng_liu@g.harvard.edu>?Subject=Your UAI 2023 paper}{Shuheng Liu}{}}
\author[2]{\href{mailto:<xh2554@columbia.edu>?Subject=Your UAI 2023 paper}{Xiyue Huang}{}}
\author[3]{\href{mailto:<pavlos@seas.harvard.edu>?Subject=Your UAI 2023 paper}{Pavlos Protopapas}{}}
\affil[1, 3]{
    Institute for Applied Computational Science\\
    Harvard University\\
    Cambridge, Massachusetts, USA
}
\affil[2]{
    Data Science Institute\\
    Columbia University\\
    New York, New York, USA
}

\begin{document}
\maketitle

\begin{abstract}
    Neural networks are universal approximators and are studied for their use in solving differential equations.
    However, a major criticism is the lack of error bounds for obtained solutions.
    This paper proposes a technique to rigorously evaluate the error bound of Physics-Informed Neural Networks (PINNs) on most linear ordinary differential equations (ODEs), certain nonlinear ODEs, and first-order linear partial differential equations (PDEs).
    The error bound is based purely on equation structure and residual information and does not depend on assumptions of how well the networks are trained.
    We propose algorithms that bound the error efficiently. 
    Some proposed algorithms provide tighter bounds than others at the cost of longer run time.
\end{abstract}

\section{Introduction}
    Differential equations (DEs) are a useful mathematical tool for describing various phenomena in natural sciences, engineering, and humanity studies. 
    As universal approximators, neural networks are powerful in approximating unknown functions. 
    With back-propagation and modern computing devices, neural networks are convenient to differentiate, making them an ideal choice for solving differential equations.

    However, a major criticism of neural network solutions to DEs is the lack of error bound. 
    Traditional numerical methods, such as the finite difference method (FDM) and the finite element method (FEM), compute numerical solutions with known error bounds.
    Unlike traditional methods, the error bounds of neural network solutions are not well-studied.
    Therefore, solving DEs with neural networks requires ad hoc customization and empirical hyperparameter finetuning.
    If the error of \textit{any} given network can be bounded, we can train neural networks until the error falls below a specified tolerance threshold.

    Our contribution is that we propose rigorous error-bounding algorithms for any neural network solution to certain classes of equations, including linear ODEs, certain nonlinear ODEs, and first-order linear PDEs.
    These algorithms can also be extended to bound the error of other classes of equations as well.
    The proposed algorithms only use residual information and equation structure as inputs and do not rely on assumptions of finetuning.

    Section \ref{section:symbols-and-notations} introduces the symbols and notations adopted in this paper.
    Section \ref{section:literature-review} reviews the literature on emerging areas of research that are relevant to solving DEs with neural networks.
    Section \ref{section:existing-work} explains the existing effort to bounding the error of neural network DE solutions.
    Sections \ref{section:error-bound-for-ode} and \ref{section:error-bound-for-pde} propose various algorithms for the error bound of ODEs and PDEs, respectively.
    Section \ref{section:experiments} uses the method of manufactured solution to verify the validity of each error-bounding algorithm and provides visualization of the tightness of the bounds.

\section{Symbols and Notations} \label{section:symbols-and-notations}
    DEs in this paper are posed w.r.t. unknown function $v$,
        \begin{equation*}
            \mathcal{D} v = f,
        \end{equation*}
    where $\mathcal{D}$ is a possibly nonlinear differential operator and $f$ is some forcing function.
    Unlike the exact solution $v(\cdot)$, a neural network solution $u(\cdot)$ does not strictly satisfy the equation.
    Instead, it incurs an additional residual term, $r$, which the network aims to minimize, to the equation, 
        \begin{equation*}
            \mathcal{D} u = f + r.
        \end{equation*}
    The input to $v$, $u$, $f$, and $r$ is time $t$ for ODEs and spatial coordinates $(x, y)$ for PDEs.
    We limit our reasoning to 2-dimensional PDEs in this work.
    In cases with multiple unknown functions, we use vector notations $\vect{v}$, $\vect{u}$, and $\vect{r}$ instead of the scalar notations $v$, $u$, and $r$.

    The loss function of the network solution is defined as the $L^2$ norm of residual $r$ over the domain of interest,
        \begin{align}
            \Loss{}(u) &:= \frac{1}{|I|} \int_{I} \|r\|^2 \mathrm{d}I = \frac{1}{|I|} \int_{I} \|\mathcal{D} u - f\|^2 \mathrm{d}I,
        \end{align}
    where a spatial domain $\Omega$ is substituted for the temporal domain $I$ in the case of a PDE.

\subsection{INITIAL AND BOUNDARY CONDITIONS}\label{section:initial-and-boundary-conditions}
    For a neural network to satisfy initial or boundary conditions, we apply a technique called \textit{parametrization}. 
    As an intuitive example, the parametrization $u(t) = (1 - e^{-t}) \Net(t) + v(0)$ guarantees that $u(t)$ satisfies the initial condition $u(0)=v(0)$ regardless of the network $\Net(\cdot)$.
    This does not affect the capability of $\Net(\cdot)$ to learn any solution.

    The parametrization is more complicated for higher-order ODEs and most PDEs and has been extensively studied by \cite{lagaris1998artificial}, \cite{lagaris2000neural}, \cite{mcfall2009artificial}, \cite{lagari2020systematic}, and \cite{sukumar2021exact}.
    In this work, we assume all initial and boundary conditions are exactly satisfied.

\subsection{ERROR AND ERROR BOUND}
    The error of a network solution $u$ is defined as 
        \begin{equation}
            \Err := u - v.
        \end{equation}
    We are interested in \textit{bounding} the error with a scalar function $\Bound$ such that 
        \begin{equation}
            \|\Err(t)\| \leq \Bound(t) \quad \text{or} \quad \|\Err(x, y)\| \leq \Bound(x, y)
        \end{equation}
    where $\|\Err\| = \|u - v\|$ is the \textit{absolute error}.
    If $\Bound$ takes on the same value $B \in \mathbb{R}^{+}$ over the domain, it can be replaced with a constant $B$.

    Notice that multiple bounds $\Bound$ exist for the same network solution $u$.
    For example, $|\Err(t)| \leq \Bound^{(1)}(t) \leq  \Bound^{(2)}(t) \leq \dots \leq B$ are bounds in decreasing order of tightness. Tighter bounds incur a higher computational cost, and looser bounds (such as constant $B$) are faster to compute.

    A summary of the applicability, restraints, run-time complexity, and relative tightness of all proposed algorithms is listed in Table \ref{tab:alogrithm-overview}.

    \begin{table*}[tbh]
        \centering
        \caption{\textbf{Overview of Proposed Algorithms.}\, The symbols in run-time analysis are defined and explained in detail in Sections \ref{section:error-bound-for-ode} and \ref{section:error-bound-for-pde}, with the exception of $K$, which is the number of steps used in each numerical integration.}
        \label{tab:alogrithm-overview}
        \begin{tabularx}{\textwidth}{ccXcc}
          \toprule 
          \bfseries Algorithm & \bfseries Applicable to & \bfseries Restraint & \bfseries Run-Time & \bfseries Comment \\
          \midrule 
          Algorithm \ref{alg:single-linear-ode-constant-coeff-loose} & Linear ODE & Semi-stable & $O(L)$ & Looser than Alg \ref{alg:single-linear-ode-constant-coeff-tight}\\
          Algorithm \ref{alg:single-linear-ode-constant-coeff-tight} & Linear ODE &  & $O(nL)$ & Tighter than Alg \ref{alg:single-linear-ode-constant-coeff-loose}\\
          Algorithm \ref{alg:system-bound} & Linear ODE System &  & $O(n^3 L)$ & Norm and elementwise bounds\\
          Algorithm \ref{alg:nonlinear-iterative} & Nonlinear ODE & Nonlinear term is $\varepsilon v^k$ & $O(JnL)$ & Bounded solution for family of DEs\\
          Algorithm \ref{alg:linear-first-order-pde-constant} & Linear 1st-Order PDE & Coeff. $c \neq 0$ over domain & $O(\text{mesh})$ & Constant bound; Looser than Alg \ref{alg:linear-first-order-pde-general}\\
          Algorithm \ref{alg:linear-first-order-pde-general} & Linear 1st-Order PDE & Solvable characteristics & $O(KL)$ & Tigher than Alg \ref{alg:linear-first-order-pde-constant} if computable\\
          \bottomrule 
        \end{tabularx}
    \end{table*}

\section{LITERATURE REVIEW} \label{section:literature-review}
    \cite{hornik1989multilayer} showed that neural networks are universal function approximators. 
    \cite{lagaris1998artificial} first studied the application of neural networks in solving DEs.
    The term \textit{physics-informed neural networks}, or PINNs, was first introduced by \cite{raissi2019physics} to name neural networks that satisfy DEs while fitting observed data points. 
    Although we train PINNs only to solve DEs without any observed data in this work, the error-bounding algorithms we propose work for any given neural network, regardless of the training process.

    \cite{flamant2020solving} and \cite{DesaiShaan2021OTLo} showed that one main advantage of neural networks over traditional numerical methods, such as FDM and FEM, is that neural networks can potentially learn the structure of the solution space and give a bundle of solutions $u(\vect{x}; \Theta)$ for different equation setup and initial/boundary conditions parameterized by $\Theta$.
    For traditional methods, a new solution must be recomputed for any slight changes in equation setup or initial/boundary conditions.

    Some effort has been made to redefine the objective loss function. 
    \cite{yu2017deep} applied the Ritz method to a particular class of variational problems.
    \cite{mattheakis2020hamiltonian} incorporated an additional constraint to force the network to learn solutions with energy conservation.
    \cite{parwani2021adversarial} used an adversarial network for sampling in particular areas of the domain where the residual is large.

    There are also works that study the failure modes of PINNs and quantify the error of PINN solutions in recent years. 
    \cite{graf2021uncertainty} worked on quantifying the uncertainty of PINNs using the Bayesian framework.
    \cite{krishnapriyan2021characterizing} characterized possible failure modes of PINNs by studying the performance of PINNs on simple problems and analyzing their loss landscape. 
    \cite{krishnapriyan2021characterizing} also concluded that optimization difficulty is the essential cause of failure.

    Our work uncovers the mathematical relationship between residual information and the error of PINNs on several classes of ODEs and PDEs. 
    We propose different algorithms for various classes of equations and experimentally validate these algorithms.

\section{EXISTING WORK}\label{section:existing-work}
    \cite{sirignano2018dgm} showed that for a class of quasi-linear parabolic PDEs, a neural network with a single hidden layer and sufficiently many hidden units could arbitrarily approximate the exact solutions.
    \cite{guo2022energy} proposed an energy-based \textit{constitutive relation error} bound for elasticity problems.

    \cite{de2022errorhyperbolic} derived an error bound for ReLU networks on parametric hyperbolic conservation laws.
    \cite{de2022errorkolmogorov} showed that there exists some PINN with arbitrarily small residual for Kolmogorov PDEs.
    \cite{de2022generic} derived an error bound for operator learning with PINNs.
    The works of \citeauthor{de2022errorhyperbolic} mentioned above did not bound the error of every given network.
    Instead, they mathematically proved the existence of a network with errors below a specified bound, under certain assumptions of network architecture, including width, depth, and activation functions. 
    The question remaining to be answered is how to overcome optimization difficulties and find such a neural network.

    Our work differs from the above in that we bound the error of \textit{any} neural network regardless of finetuning, even networks with randomly initialized weights.
    Our algorithms only depend on inputs of residual information $r$, often used as training loss, and equations structure $\mathcal{D} v = f$.
    The output is a (possibly constant) function that guarantees to bound the error at any point in domain.

\section{ERROR BOUND FOR ODE}  \label{section:error-bound-for-ode}
    This section considers both linear and nonlinear ODEs over the temporal domain $I=[0, T]$. 
    Initial conditions are imposed on $\frac{\mathrm{d}^k}{\mathrm{d}t^k}v(t=0)$ for $k = 0, \dots, (n - 1)$, where $n$ is the highest order of derivative terms in ODE.

\subsection{ERROR BOUND FOR LINEAR ODE}\label{section:error-bound-for-linear-odes}
    Consider the linear ODE $\L v(t) = f(t)$, where $\L$ is a linear differential operator. 
    Its neural network solution $u$ satisfies $\L u(t) = f(t) + r(t)$. 
    Since error $\Err := u - v$, there is
        \begin{equation} \label{eq:linear-error-master}
            \L \Err(t) = r(t).
        \end{equation}
    With the assumption in Section \ref{section:initial-and-boundary-conditions} that $u$ satisfies the initial conditions at $t=0$, there is
        \begin{equation} \label{eq:linear-error-initial-condition}
            \Err(0) = 0, \quad \dt{}{}\Err(0) = 0, \quad \dnt{2}{}\Err(0) = 0, \quad \dots 
        \end{equation}
    With initial conditions \ref{eq:linear-error-initial-condition} known, a unique inverse transform $\L^{-1}$to $\L$ exists. 
    Applying $\L^{-1}$ to Eq. \ref{eq:linear-error-master}, there is 
        \begin{equation}\label{eq:linear-error-inverse-master}
            \Err(t) = \L^{-1} r(t).
        \end{equation}
    Hence, bounding the absolute error $\left|\Err\right|$ is equivalent to bounding $\left|\L^{-1} r\right|$. 
    Notice that only a) the equation structure $\L$ and b) the residual information $r$ are relevant to estimating the error bound. 
    All other factors, including parameters of the neural network $u$, forcing function $f$, and initial conditions, do not affect the error bound at all.

\subsubsection{Single Linear ODE with Constant Coefficients}\label{section:single-linear-ode-with-constant-coefficients}
    Consider the case where $\L = \frac{\mathrm{d}^n}{\mathrm{d}t^n} + \sum_{j=0}^{n - 1} a_j \frac{\mathrm{d}^j}{\mathrm{d}t^j}$ consists of only constant coefficients $a_0, a_1, \dots, \in \mathbb{R}$.
    Its characteristic polynomial (defined below) can be factorized into
        \begin{equation} \label{eq:single-linear-ode-characteristic-polynomial-factorization}
            \lambda^n + a_{n-1}\lambda^{n-1} + \dots + a_0 = \prod_{j=1}^{n}(\lambda - \lambda_j),
        \end{equation}
    where $\lambda_1, \dots, \lambda_n \in \mathbb{C}$ are the characteristic roots. 

    It can be shown that, for a semi-stable system ($\Re{\lambda_j} \leq 0$ for all $\lambda_j$), an error bound can be formulated as
    \begin{equation} \label{eq:linear-ode-const-loose-bound}
        \left|\Err(t)\right| \leq \Bound_{loose}(t) := C_{\lambda_{1:n}}\, R_{\max}\, t^{Z},
    \end{equation}
    where $0\leq Z \leq n$ is the number of $\lambda_j$ whose real part is $0$, $C_{\lambda_{1:n}} := \frac{1}{Z!}\prod_{j=1; \lambda_j\neq 0}^{n} \frac{1}{\Re{-\lambda_j}}$ is a constant coefficient, and $R_{\max}:=\max_{t\in I} |r(t)|$ is the maximum absolute residual. 
    Knowing bound \ref{eq:linear-ode-const-loose-bound} is sufficient to qualitatively estimate the error for applications where only the order of error is concerned. See Alg. \ref{alg:single-linear-ode-constant-coeff-loose} for reference.

    \begin{algorithm}
        \caption{Loose Error Bound Estimation for Linear ODE with Constant Coefficients\quad (Requires Semi-Stability)}\label{alg:single-linear-ode-constant-coeff-loose}
        \textbf{Input:} Coefficients $\left\{a_j\right\}_{j=0}^{n-1}$ for operator $\L$, residual information $r(\cdot)$, domain of interest $I = [0, T]$, and a sequence of time points $\left\{t_\ell\right\}_{\ell=1}^{L}$ where error bound is to be evaluated\\
        \textbf{Output:} Error bound at given time points $\left\{\Bound(t_\ell)\right\}_{\ell=1}^{L}$
        \begin{algorithmic}
            \Require $\L$ is semi-stable, and $t_\ell \in I$ for all $\ell$
            \Ensure $\left|\Err(t_\ell)\right| \leq \Bound(t_\ell)$ for all $\ell$
            \State $\{\lambda_j\}_{j=1}^{n} \gets$ numerical roots of $\lambda^n+a_{n-1}\lambda^{n-1}+\dots=0$ 
            \State \textbf{assert} $\lambda_j \leq 0$ for $1 \leq j \leq n$ 
            \State $Z, C \gets 0, 1$
            \For{$j\gets 1\dots n$}
                \If{$\Re{\lambda_j} = 0$}
                    \State $Z \gets Z + 1$
                \Else
                    \State $C \gets C / \Re{-\lambda_j}$
                \EndIf
            \EndFor
            \State $R_{\max} \gets \max_{\tau \in I} |r(\tau)|$ \Comment{Use linspace with mini-steps}
            \State $\left\{\Bound(t_\ell)\right\}_{\ell=1}^{L} \gets \left\{\frac{C}{Z!}R_{\max}\, t_\ell^{Z}\right\}_{\ell=1}^{L}$
            \State \textbf{return} $\left\{\Bound(t_\ell)\right\}_{\ell=1}^{L}$
        \end{algorithmic}
        \textbf{Note}: \cite{jenkins1970three} solves polynomial roots.
    \end{algorithm}

    An issue with Eq. \ref{eq:linear-ode-const-loose-bound} and Alg. \ref{alg:single-linear-ode-constant-coeff-loose} is that they assume $\Re{\lambda_j} \leq 0$ for all characteristic roots $\lambda_j$. 
    To address this issue, we propose an alternative error-bounding Alg. \ref{alg:single-linear-ode-constant-coeff-tight}, which requires more computation but does not require the system to be semi-stable and provides a tighter bound.

    Notice that the bounds of $\Err$ in Eq. \ref{eq:linear-error-inverse-master} can be estimated if the inverse operator $\L^{-1}$ is known. 
    Let Eq. \ref{eq:single-linear-ode-characteristic-polynomial-factorization} be the factorization of characteristic polynomial of $\L$.
    Define operator $\I_{\lambda}$ as \footnote{
        This paper assumes the network solution exactly satisfies the initial conditions as discussed in Section \ref{section:initial-and-boundary-conditions}. 
        However, all our algorithms can be extended to cases where the network solution differs from the exact solution by some value. 
        This is achieved by replacing $\I_\lambda \psi(t)$ with $\I_{\lambda, \delta} \psi(t) = \I_\lambda \psi(t) + \delta e^{\lambda t}$ in Eq. \ref{eq:integral-operator-definition}. 
    }
    \begin{equation} \label{eq:integral-operator-definition}
        \I_\lambda \psi(t) := e^{{\lambda} t} \int_{0}^{t} e^{-{\lambda} \tau} \psi(\tau) \mathrm{d}\tau, \quad \forall \psi : I \to \mathbb{C}.
    \end{equation}
    We show in supplementary material that $\L^{-1} = \I_{\lambda_{n}} \circ \I_{\lambda_{n-1}} \circ \dots \circ \I_{\lambda_1}$ and that $\left|\I_{\lambda} \psi\right| \ \leq \I_{\Re{\lambda}} |\psi|$ for any $\lambda \in \mathbb{C}$ and function $\psi$.
    Hence, another error bound can be formulated as
    \begin{equation} \label{eq:single-linear-ode-inverse-operator-inequality}
        \Bound_{tight}(t) := \left(\I_{\Re{\lambda_{n}}} \circ \dots \circ \I_{\Re{\lambda_1}}\right) |r(t)|.
    \end{equation}
    It is also proven in supplementary material that $\Bound_{tight}$ is tighter than $\Bound_{loose}$ when $\Bound_{loose}$ is applicable,
    \begin{equation} \label{eq:single-linear-ode-tight-and-loose}
        \left|\Err(t)\right| \leq \Bound_{tight}(t) \leq \Bound_{loose}(t) \quad \forall t \in I.
    \end{equation}
    Based on Eq. \ref{eq:single-linear-ode-inverse-operator-inequality}, we propose Alg. \ref{alg:single-linear-ode-constant-coeff-tight}, which computes $\Bound_{tight}$ by repeatedly evaluating integrals in \ref{eq:integral-operator-definition} using the cumulative trapezoidal rule.

    \begin{algorithm}
        \caption{Tighter Error Bound Estimation for Linear ODE with Constant Coefficients\quad  (Stable and Unstable)}\label{alg:single-linear-ode-constant-coeff-tight}
        \textbf{Input \& Output:} Same as Alg. \ref{alg:single-linear-ode-constant-coeff-loose}
        \begin{algorithmic}
            \Require Same as Alg. \ref{alg:single-linear-ode-constant-coeff-loose}, except $\L$ can be unstable
            \Ensure Same as Alg. \ref{alg:single-linear-ode-constant-coeff-loose}
            \State $\{\lambda_j\}_{j=1}^{n} \gets$ numerical roots of $\lambda^n+a_{n-1}\lambda^{n-1}+\dots=0$
            \State $\left\{t_k\right\}_{k=0}^{K} \gets$ linspace($0$, $T$, \normalfont{sufficient steps})
            \State $\left\{\Bound(t_k)\right\}_{k=0}^{K} \gets \left\{|r(t_k)|\right\}_{k=0}^{K}$
            \For{$j \gets 1 \dots n$}
                \State intgr$_{k=0}^{K} \gets$ CumTrap($\left\{e^{-\lambda_j t_{k}} \Bound(t_k)\right\}_{k=0}^{K}$, $\left\{t_k\right\}_{k=0}^{K}$) 
                \State $\left\{\Bound(t_k)\right\}_{k=0}^{K} \gets \left\{e^{\lambda_j t_{k}}\cdot \text{intgr}_k \right\}_{k=0}^{K}$ 
            \EndFor
            \State $\left\{\Bound(t_\ell)\right\}_{\ell=1}^{L} \gets $ Interp($\left\{\Bound(t_k)\right\}_{k=0}^{K}$, $\left\{t_k\right\}_{k=0}^{K}$, $\left\{t_\ell\right\}_{\ell=0}^{L}$) 
            \State \textbf{return} $\left\{\Bound(t_\ell)\right\}_{\ell=1}^{L}$ 
        \end{algorithmic}

        \textbf{Note}: CumTrap($\{y_k\}_{k=1}^K$, $\{x_k\}_{k=1}^K$) computes cumulative integral $\int_{0}^x y(x)\mathrm{d}x$ at discrete points $\{x_k\}_{k=1}^K$ using trapezoidal rule.\\
        \textbf{Note}: Interp($\{y_k\}_{k=1}^K$, $\{x_k\}_{k=1}^K$, $\{x_\ell\}_{\ell=1}^L$) computes interpolant to a function with given discrete data points $\{(x_k, y_k)\}_{k=1}^K$ evaluated at $\{x_\ell\}_{\ell=1}^L$.
    \end{algorithm}

    Before moving on to the next section, we discuss the effect of numerical error in Alg. \ref{alg:single-linear-ode-constant-coeff-tight} and all pursuant algorithms which involve numerical integration.
    Empirically, the error introduced by numerical integration in negligible for most cases.
    To ensure the accuracy of the error bound, we recommend using a slightly modified trapezoidal rule that, instead of directly using the node value,  takes the maximum of the node and its two adjacent nodes.
    This modification leads to a slighltly looser bound.
    Yet, this influence is almost negligible in practice.

\subsubsection{Single Linear ODE of the General Form}
   In general, the coefficients for $\L$ can be functions of $t$. Namely, $\L = \frac{\mathrm d^n}{\mathrm d t^n} +  \sum_{j=0}^{n-1}a_j(t)\frac{\mathrm d^j}{\mathrm d t^j}$.
   Similar to Eq. \ref{eq:single-linear-ode-characteristic-polynomial-factorization}, $\L$ have characteristic roots $\{\lambda_{j}(t)\}_{j=1}^{n}$ as functions of $t$,
        \begin{equation*} \label{eq:functional-factorization}
            \lambda^n + a_{n-1}(t)\lambda^{n-1} + \dots + a_0(t) = \prod_{j=1}^{n}(\lambda - \lambda_j(t)).
        \end{equation*}
    We can replace constant $\lambda_j$ with functions $\lambda_j(t)$ in Eq. \ref{eq:integral-operator-definition} and compute bound $\Bound_{tight}$ as in Eq. \ref{eq:single-linear-ode-inverse-operator-inequality}.
    However, the factorization in Eq. \ref{eq:functional-factorization} is hard to implement in practice except for the first-order case  where $\L v = \frac{\mathrm{d}v}{\mathrm{d}t} + a_0(t)v$. 
    Cases of second order and higher are out of the scope of this paper.

\subsubsection{System of Linear ODEs with Constant Coefficients} \label{section:system-of-linear-odes-with-constant-coefficients}
    Consider a system of linear ODEs with constant coefficients 
        \begin{equation}\label{eq:linear-system-master}
            \frac{\mathrm{d}}{\mathrm{d}t}\vect{v} + A\vect{v} = \vect{f}(t)
        \end{equation}
    where $\vect{v}$ and $\vect{f}$ are $\mathbb{R}^n$ vectors and $A$ is a $n\times n$ matrix. Denote the Jordan canonical form of $A$ as,
        \begingroup 
        \setlength\arraycolsep{1pt}
        \begin{equation}\label{eq:jordan-definition}
            J = P^{-1}AP= \begin{pmatrix}
                J_1 & & \\[-0.25em]
                & \ddots & \\[-0.25em]
                & & J_K
            \end{pmatrix}
            {\text{ where }}
            J_k = \begin{pmatrix}
                \lambda_k & 1\\[-0.75em]
                & \lambda_k & \ddots\\[-0.75em]
                & & \ddots & 1\\[-0.25em]
                & & & \lambda_k
            \end{pmatrix}.
        \end{equation}
        \endgroup
    Let $n_k$ be the size of Jordan block $J_k$, we construct an operator matrix $\pmb{\I} = \text{diag}(\vect{I}_1, \vect{I}_2, \dots)$, where 
        \begingroup 
        \setlength\arraycolsep{1pt}
        \begin{equation}\label{eq:operator-block}
            \vect{I}_k = \begin{pmatrix}
                \I_{-\Re{\lambda_k}} & \I_{-\Re{\lambda_k}}^2 & \dots &\I_{-\Re{\lambda_k}}^{n_k} \\[1ex]
                0 & \I_{-\Re{\lambda_k}} & \dots &\I_{-\Re{\lambda_k}}^{n_k-1} \\[-1ex]
                \vdots & \vdots & \ddots & \vdots \\
                0 & 0 & \dots & \I_{-\Re{\lambda_k}}
            \end{pmatrix}.
        \end{equation}
        \endgroup
    An \textit{elementwise bound} (vector) $\pmb{\Bound}(t)$ can be formulated as 
        \begin{equation}\label{eq:system-component-bound}
            \pmb{\Err}^{\abs}(t) \preceq \pmb{\Bound}(t) := P^{\abs} \pmb{\I}\left[(P^{-1})^{\abs} \ \vect{r}^{\abs}\right](t),
        \end{equation}
    where superscript $\abs$ denotes taking elementwise absolute value and symbol $\preceq$ denotes elementwise inequality. In the meantime, a \textit{norm bound} (scalar) $\Bound(t)$ also exists
        \begin{equation}\label{eq:system-norm-bound}
            \left\|\pmb{\Err}(t)\right\| \leq \Bound(t) := \mathrm{cond}(P)\left\|\pmb{\I}\big[\|\vect{r}\|\vect{1}\big](t)\right\|
        \end{equation}
    where $\mathrm{cond}(P)$ is the conditional number of $P$ w.r.t. induced matrix norm and $\vect{1}$ is an $n\times 1$ column vector of $1$s. Proof of Eq. \ref{eq:system-component-bound} and Eq. \ref{eq:system-norm-bound} can be found in in supplementary material.
    See Alg. \ref{alg:system-bound} for implementation.

    \begin{algorithm}
        \caption{ODE System Bound (norm and elementwise)}\label{alg:system-bound}
        \textbf{Input:} Coefficient matrix $A \in \mathbb{R}^{n\times n}$, residual vector $\vect{r}(t)$, and a sequence of points $\left\{t_\ell\right\}_{\ell=1}^{L}$ where error is to be bounded\\
        \textbf{Output:} Norm bound (scalar) $\left\{\Bound(t_\ell)\right\}_{\ell=1}^{L}$ and componentwise bound (vector) $\left\{\pmb{\Bound}(t_\ell)\right\}_{\ell=1}^{L}$ at given time points
        \begin{algorithmic}
            \Ensure $\|\Err(t_\ell)\| \leq \Bound(t_\ell)$ and $\Err(t_\ell) \preceq \pmb{\Bound}(t_\ell)$ for all $\ell$

            \State $J, P \gets $ Jordan canonicalization of $A = PJP^{-1}$
            \For{each Jordan block $J_k$ of shape $n_k \times n_k$}
                \State $\vect{I}_k \gets$ construct operator block using Eq. \ref{eq:operator-block} 
            \EndFor
            \State $\pmb{\I} \gets$ diag($\vect{I}_1$, $\vect{I}_2$,  \dots)
            \State $\left\{\pmb{\Bound}(t_\ell)\right\}_{\ell=1}^{L} \gets \{P^{\abs} \pmb{\I}\big[(P^{-1})^{\abs} \vect{r}^{\abs}\big](t_\ell)\}_{\ell=1}^{L}$
            \State $\left\{\Bound(t_\ell)\right\}_{\ell=1}^{L} \gets \{\mathrm{cond}(P)\left\|\pmb{\I}\big[\|\vect{r}\|\vect{1}\big](t_\ell)\right\|\}_{\ell=1}^{L}$
            \State \textbf{return} $\left\{\Bound(t_\ell)\right\}_{\ell=1}^{L}$, $\left\{\pmb{\Bound}(t_\ell)\right\}_{\ell=1}^{L}$
        \end{algorithmic}
    \end{algorithm}

\subsection{NONLINEAR ODE}
    Nonlinear ODEs are hard to solve in general. 
    In this work, we only deal with nonlinear ODEs with a single nonlinear term of the form $\varepsilon v^k(t)$, where $\varepsilon \in \mathbb{R}$ is a small number.
    Ideally, $|\varepsilon| \ll 1$. 
    The exact requirement for $\epsilon$ is given in Section \ref{section:expansion-of-bounds}.
    The value of $\varepsilon$ can vary within a certain range or be fixed.
    With the perturbation technique, we obtain a family of solutions $v(t;\varepsilon)$ parameterized by $\varepsilon$ at the cost of solving a (countable) collection of equations. 
    As explained below in Section \ref{section:perturbation-theory}, we train finitely many networks, each approximately solving an equation in the collection.

\subsubsection{Perturbation Theory} \label{section:perturbation-theory}
    Consider the nonlinear ODE with nonlinear term $\varepsilon v^k(t)$,
        \begin{equation} \label{eq:nonlinear-ode-master}
            \L v(t) + \varepsilon v^k(t) = f(t),
        \end{equation}
    where $\L$ is a linear differential operator discussed in \ref{section:error-bound-for-linear-odes} and initial conditions are specified for the system at time $t=0$. 
    Notice that each $\varepsilon \in \mathbb{R}$ corresponds to a solution $v(t; \varepsilon)$. 
    We expand the solution $v(t; \varepsilon)$ in terms of $\varepsilon$
        \begin{equation} \label{eq:nonlinear-solution-expansion}
            v(t; \varepsilon) = \sum_{j=0}^{\infty} \varepsilon^j v_j(t) = v_0(t) + \varepsilon v_1(t) + \dots
        \end{equation}
    Only $v_0(t)$ is subject to the original initial conditions at $t=0$, while other components, $v_1$, $v_2$, \dots, have initial conditions of $0$ at $t=0$.
    Substituting Eq. \ref{eq:nonlinear-solution-expansion} into Eq. \ref{eq:nonlinear-ode-master},
        \begin{gather}
            \L \sum_{j=0}^{\infty} \varepsilon^j v_j + \varepsilon \left(\sum_{j=0}^{\infty} \varepsilon^j v_j\right)^k = f \\
            \sum_{j=0}^{\infty} \varepsilon^j \L v_j + \sum_{j=0}^{\infty} \varepsilon^{j+1} \sum_{\substack{j_1+\dots+j_k = j\\j_1, \dots, j_k \geq 0}}v_{j_1}\dots v_{j_k} = f \\[-0.5em]
            \L v_0 + \sum_{j=1}^{\infty} \varepsilon^j \Bigg(\L v_j + \sum_{\substack{j_1+\dots+j_k = j - 1\\j_1, \dots, j_k \geq 0}}v_{j_1}\dots v_{j_k}\Bigg)= f \label{eq:nonlinear-equation-expansion}.
        \end{gather}
    In order for Eq. \ref{eq:nonlinear-equation-expansion} to hold true for all $\varepsilon$, the coefficients for each $\varepsilon^j$ must match on both sides of Eq. \ref{eq:nonlinear-equation-expansion}. Hence,
        \begin{alignat}{6}
            &\L v_0 &&= f \label{eq:expansion-epsilon-0}\\
            &\L v_1 + v_0^k &&= 0 \label{eq:expansion-epsilon-1}\\
            &\L v_2 + k v_0^{k-1}v_1 &&= 0 \label{eq:expansion-epsilon-2} \\
            &\L v_3 + \frac{k(k-1)}{2} v_0^{k-2}v_1^2 + k v_0^{k-1}v_2 &&= 0 \label{eq:expansion-epsilon-3} \\[-1em]
            &\vdots &&\phantom{=}\,\,\,\,\vdots\nonumber
        \end{alignat}

    For $\varepsilon = 0$, Eq. \ref{eq:nonlinear-solution-expansion} is reduced to $v_0(t)$, which solves the linear problem $\L v=f$. 

    \begingroup
        \setlength{\itemsep}{0pt}
        \setlength{\parskip}{0pt}
        The above system can be solved in a \textit{sequential} manner, either analytically or using neural networks,
        \begin{enumerate}
            \item Eq. \ref{eq:expansion-epsilon-0} is linear in $v_0$ and can be solved first. 
            \item With $v_0$ known, Eq. \ref{eq:expansion-epsilon-1} is linear in $v_1$ and can be solved for $v_1$. 
            \item Similarly, with $v_0$ and $v_1$ known, Eq. \ref{eq:expansion-epsilon-2} is linear in $v_2$ and can be solved for $v_2$.
            \item The process can be repeated for Eq. \ref{eq:expansion-epsilon-3} and beyond. Only a linear ODE is solved each time.
        \end{enumerate}
        To solve the system with PINNs, we approximate exact solutions $\left\{v_j(t)\right\}_{j=1}^{\infty}$ with neural network solutions $\left\{u_j(t)\right\}_{j=0}^{J}$ trained sequentially on Eq. \ref{eq:expansion-epsilon-0}, Eq. \ref{eq:expansion-epsilon-1}, and beyond. 
        In practice, we only consider components up to order $J$ to avoid the infinity in expansion \ref{eq:nonlinear-solution-expansion}. 
        Ideally, $J$ should be large enough so that higher order residuals in expansion \ref{eq:nonlinear-solution-expansion} can be neglected.
    \endgroup

    After obtaining $\left\{u_j(t)\right\}_{j=0}^{J}$, we can reconstruct the solution $u(t;\varepsilon) = \sum_{j=0}^{J} \varepsilon^j u_j(t)$ to the original nonlinear equation \ref{eq:nonlinear-ode-master} for varying $\varepsilon$.
    See Alg. \ref{alg:nonlinear-iterative} for details.

\subsubsection{Expansion of Bounds}\label{section:expansion-of-bounds}
    The absolute error $|\Err(t;\varepsilon)| = |u(t;\varepsilon) - v(t;\varepsilon)|$is given by 
        \begin{align}
            |\Err(t; \varepsilon)| 
            &= \left|\sum_{j=0}^{J} \varepsilon^{j} \Big(u_j(t) - v_j(t)\Big) - \sum_{j=J+1}^{\infty} \varepsilon^j v_j(t)\right| \nonumber \\[-0.5em]
            &\leq \sum_{j=0}^{J} \Big|\Err_{j}(t)\Big||\varepsilon|^j + \left|\sum_{j=J+1}^{\infty}\varepsilon^j v_j(t)\right| 
        \end{align}
    where $\Err_{j}(t) := u_j(t) - v_j(t)$ is the \textit{component error} between $u_j(t)$ and $v_j(t)$.
    Let $\Bound_{j}$ denote the \textit{bound component} such that $|\Err_{j}(t)| \leq \Bound_j(t)$.
    Assuming $J$ is large and $\varepsilon$ is small such that higher order terms $\left|\sum_{j=J+1}^{\infty}\varepsilon^j v_j(t)\right|$ are negligible, there is 
        \begin{equation} \label{eq:nonlinear-bound-components}
            \Big|\Err(t; \varepsilon)\Big| \leq \Bound(t; \varepsilon) := \sum_{j=0}^{J} \Bound_j(t)\,|\varepsilon|^j 
        \end{equation}
    where each bound component $\Bound_j$ can be evaluated using the techinque in Section \ref{section:error-bound-for-linear-odes}. 
    See Alg. \ref{alg:nonlinear-iterative} for details.

    \begin{algorithm}
        \caption{Iterative Method for Solution and Error Bound of Nonlinear ODE \ref{eq:nonlinear-ode-master}} \label{alg:nonlinear-iterative}
        \textbf{Input:} Linear operator $\L$, nonlinear degree $k$, domain $I=[0, T]$, highest order $J$ for expansion, and a sequence $\left\{(t_\ell, \varepsilon_\ell)\right\}_{\ell=1}^{L}$ where solution $u(t; \varepsilon)$ and error bound $\Bound(t; \varepsilon)$ are to be evaluated \\
        \textbf{Output:} Solution $\left\{u(t_\ell; \varepsilon_\ell)\right\}_{\ell=1}^{L}$ and error bound $\left\{\Bound(t_\ell; \varepsilon_\ell)\right\}_{\ell=1}^{L}$ 
        \begin{algorithmic}
            \Require $t_\ell \in I$, and $|\varepsilon_\ell|$ to be small (ideally $|\varepsilon_\ell| \ll 1$)
            \Ensure $\Err(t_\ell; \varepsilon_\ell) \leq \Bound(t_\ell; \varepsilon_\ell)$ 

            \State $u_0, r_0, \gets$ net solution, residual of $\L u_0 = f$
            \State $\left\{\Bound_{0}(t_\ell)\right\}_{\ell=1}^L \gets$ bound of $\left|\L^{-1}r_0\right|$ at $\left\{t_\ell\right\}_{\ell=1}^L$
            \For{$j \gets 1 \dots J$} 
                \State Macro $\text{NL}_j[\phi] \gets \sum_{\substack{j_1 + \dots + j_k = j-1\\ j_1, \dots, j_k \geq 0}} \phi_{j_1} \dots \phi_{j_k}$
                \State $u_j, r_j \gets$ net solution, residual of $\L u_j + \text{NL}_j[u] = 0$
                \State $\Bound_{\text{NL}} \gets \text{upper bound of }|\text{NL}_j[u] - \text{NL}_j[v]|$
                \State $\left\{\Bound_{j}(t_\ell)\right\}_{\ell=1}^L \gets$  bound of $|\L^{-1}r_j|$+$|\L^{-1}\Bound_{\text{NL}}|$ 
            \EndFor
            \State $\left\{u(t_\ell; \varepsilon_\ell)\right\}_{\ell=1}^L \gets \big\{\sum_{j=0}^{J}\varepsilon_\ell^j u_j(t_\ell)\big\}_{\ell=1}^L $ 
            \State $\left\{\Bound(t_\ell; \varepsilon_\ell)\right\}_{\ell=1}^L \gets \big\{\sum_{j=0}^{J}\varepsilon_\ell^j \Bound_j(t_\ell)\big\}_{\ell=1}^L $ 
            \State \textbf{return} $\left\{u(t_\ell; \varepsilon_\ell)\right\}_{\ell=1}^L, \left\{\Bound(t_\ell; \varepsilon_\ell)\right\}_{\ell=1}^L$
        \end{algorithmic}

        \textbf{Note} 1: $\Bound_0$ and $\Bound_{1:J}$ can be evaluated using Alg. \ref{alg:single-linear-ode-constant-coeff-loose} or \ref{alg:single-linear-ode-constant-coeff-tight}.\\
        \textbf{Note} 2: $\Bound_\text{NL}$ can be estimated even though exact solutions $v_{0:j-1}$ are unknown. This is because $v_i \in [u_i - \Bound_i, u_i+\Bound_i]$ for all $i$, and $u_{0:j-1}$, $\Bound_{0:j-1}$ are known from previous iterations.
    \end{algorithm}

\section{ERROR BOUND FOR PDE} \label{section:error-bound-for-pde}
    This section considers first-order linear PDEs defined on a 2-dimensional spatial domain $\Omega$,\footnote{Similar techniques can be used for other classes of PDEs and higher dimensions where the method of characteristics applies.} 
        \begin{equation}\label{eq:pde-master}
            a(x, y) \px{v} + b(x, y) \py{v} + c(x, y)v = f(x, y)
        \end{equation}
    with Dirichlet boundary constraints defined on $\Gamma \subset \partial \Omega$,
        \begin{equation}\label{eq:pde-bc-master}
            v\big|_{(x, y) \in \Gamma} = g(x, y).
        \end{equation}

    We partition the domain into infinitely many characteristic curves $\mathcal{C}$, each passing through a point $(x_0, y_0) \in \Gamma$. The resulting curve is a parameterized integral curve 
        \begin{equation*} 
            \mathcal{C}: \begin{cases*}
                x'(s) = a(x, y) \\
                y'(s) = b(x, y) 
            \end{cases*} 
            \,
            \text{where}
            \,
            (\cdot)' = \ds{}
            \,\,
            \text{and} 
            \,\,
            \begin{aligned}
                x(0) &= x_0 \\
                y(0) &= y_0.
            \end{aligned}
        \end{equation*}
    For any $(x(s), y(s))$ on $\mathcal{C}$, functions $(v, a, b, c, f)$ can be viewed as univariate functions of $s$. By chain rule, there is
        \begin{equation*}
            a(x, y)\px{v} + b(x, y)\py{v} = x'(s)\px{v}  + y'(s)\py{v} = v'(s).
        \end{equation*}
    Hence, Eq. \ref{eq:pde-master} is reformulated as an ODE along curve $\mathcal{C}$,
        \begin{equation}
            v'(s) + c(s) v(s) = f(s) \quad \text{s.t. } v(0) = g(x_0, y_0),
        \end{equation}
    where $v(s)$, $c(s)$, and $f(s)$ are shorthand notations for $v(x(s),y(s))$, $c(x(s),y(s))$, and $f(x(s),y(s))$, respectively.

    In particular, if $c(x, y) \neq 0$ for all $(x, y) \in \Omega$, both sides of Eq. \ref{eq:pde-master} can be divided by $c(x, y)$, resulting in a residual of $r(x, y)/c(x, y)$ where $r(x, y)$ is the residual of the original problem. 
    By Eq. \ref{eq:linear-ode-const-loose-bound}, a constant error bound on $\mathcal{C}$ is $|\Err(s)| \leq \max_{s}\left|r(s)/c(s)\right|$. 
    Hence, a (loose) constant error bound $B$ (see Alg. \ref{alg:linear-first-order-pde-constant}) over the entire domain $\Omega$ is
        \begin{equation}
            |\Err(x, y)| \leq B :=\max_{(x, y)\in \Omega}\left|\frac{r(x, y)}{c(x, y)}\right|.
        \end{equation}

    \begin{algorithm}
        \caption{Constant Err Bound for Linear 1st-Order PDE}\label{alg:linear-first-order-pde-constant}
        \textbf{Input:} Coefficient $c(x, y)$ in Eq. \ref{eq:pde-master}, residual information $r(x, y)$ and domain of interest $\Omega$\\
        \textbf{Output:} A constant error bound $B \in \mathbb{R}^+$
        \begin{algorithmic}
            \Require $c(x, y) \neq 0$ for all $(x, y) \in \Omega$
            \Ensure $|\Err(x, y)| \leq B$ for all $(x, y) \in \Omega$

            \State $\left\{(x_k, y_k)\right\}_{k} \gets$ sufficiently dense mesh grid over $\Omega$
            \State $\displaystyle B \gets \max_{k} \left| \frac{r(x_k, y_k)}{c(x_k, y_k)}\right|$
            \State \textbf{return} $B$
        \end{algorithmic}
    \end{algorithm}

    Independent of the assumption $c(x, y)\neq 0$, in scenarios where the curve $\mathcal{C}$ passing through any $(x, y)$ can be computed, the error can be computed using Alg. \ref{alg:linear-first-order-pde-general}.

    \begin{algorithm}
        \caption{General Err Bound for Linear 1st-Order PDE}\label{alg:linear-first-order-pde-general}
        \textbf{Input:} Coefficients $a(x, y)$, $b(x, y)$, $c(x, y)$ in Eq. \ref{eq:pde-master}, residual information $r(x, y)$, domain of interest $\Omega$, Dirichlet boundary $\Gamma\subset \partial \Omega$, and a sequence of points $\left\{(x_\ell, y_\ell)\right\}_{\ell=1}^{L}$ where error is to be bounded\\
        \textbf{Output:} Error bound $\left\{\Bound(x_\ell, y_\ell)\right\}_{\ell=1}^{L}$ at given points
        \begin{algorithmic}
            \Require Integral curve of vector field $\big[a(x, y)\, b(x, y)\big]^T$ passing through any point $(x_\ell, y_\ell) \in \Omega$ is solvable
            \Ensure $|\Err(x_\ell, y_\ell)| \leq \Bound(x_\ell, y_\ell)$ for all $\ell$

            \State $\mathcal{C}_{\text{g}} \gets $ general solution to {$\begin{cases}x'(s) = a(x, y) \\ y'(s) = b(x, y)\end{cases}$}
            \For{$\ell \gets 1 \dots L$}
                \State $x(s), y(s)\gets$ instance of $\mathcal{C}_{\text{g}}$ passing through $(x_\ell, y_\ell)$
                \State $s^* \gets$ solution to $x(s) = x_\ell,\, y(s)=y_\ell$
                \State $\displaystyle \Bound(x_\ell, y_\ell) \gets e^{c(s^*)}\int_{0}^{s^*}r(x(s), y(s)) e^{-c(x(s),y(s))\,s}\mathrm{d}s$ 
            \EndFor
            \State \textbf{return} $\left\{\Bound(x_\ell, y_\ell)\right\}_{\ell=1}^{L}$
        \end{algorithmic}
    \end{algorithm}

\section{RELEVANT EXPERIMENTS}\label{section:experiments}
    In this section, we perform experiments on equations with manufactured solutions using the \textit{NeuroDiffEq} library \citep{chen2020neurodiffeq}, which provides convenient tools for training PINNs. 

    First, we train networks to solve equations and collect their residual information $r$.
    Then, we apply Alg. \ref{alg:single-linear-ode-constant-coeff-loose}--\ref{alg:linear-first-order-pde-general} (where applicable) to derive error bounds using only residual information $r$ and equation structure, characterized by its differential operator $\mathcal {D}$. 
    Lastly, we show that the absolute error strictly falls within the bounds, regardless of how well the networks are trained.

    Throughout this section, we always use networks with two hidden layers, each consisting of 32 hidden units.
    Depending on whether the problem is an ODE or PDE, a network can have a single input $t$ or two inputs $(x, y)$, but always have a single output.
    The activation function is $\tanh$. 
    Unless otherwise noted, the training domain is $I=[0, 1]$ for ODEs and $\Omega=[0,1]^2$ for PDEs. 
    We use a \textit{PyTorch} Adam optimizer with default hyperparameters to train networks for 1000 epochs.

    Notice that we list these configurations only for the reproducibility of visualizations. 
    Our error-bounding algorithm works under any other configurations.

\subsection{SINGLE LINEAR ODE WITH CONSTANT COEFFICIENTS}
    Here, we study three equations $v'' + 3v' + 2v = f(t)$, $v'' + v = g(t)$, and $v'' - v' = h(t)$, whose characteristic roots are $\{-1, -2\}$, $\{\pm i\}$, and $\{0, 1\}$ respectively. 
    By Section \ref{section:single-linear-ode-with-constant-coefficients}, the first two equations can be bounded with either Alg. \ref{alg:single-linear-ode-constant-coeff-loose} or Alg. \ref{alg:single-linear-ode-constant-coeff-tight}, while the last must be bounded with Alg. \ref{alg:single-linear-ode-constant-coeff-tight}.

\begin{figure}[!ht]
        \centering
        \includegraphics[width=\linewidth]{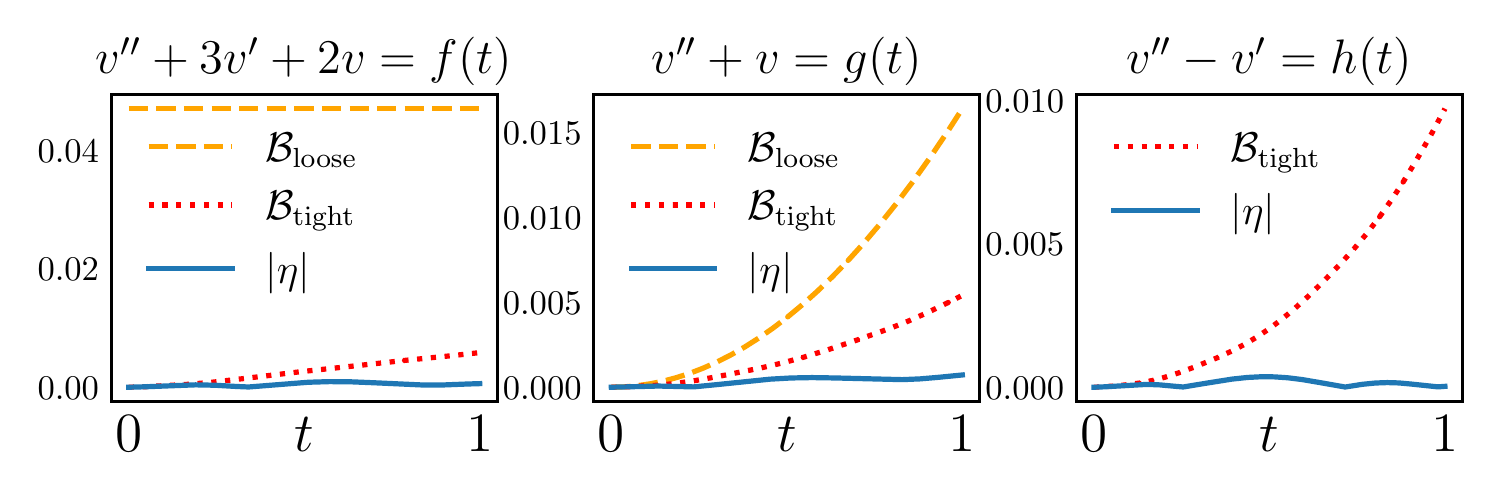}
        \caption{
            Loose bound (Alg. \ref{alg:single-linear-ode-constant-coeff-loose}) and tight bound (Alg. \ref{alg:single-linear-ode-constant-coeff-tight}) for 3 second-order linear ODE with constant coefficients.
            Notice that the loose bound cannot be applied to the third equation since it has characteristic roots with positive real part.
        }\label{fig:2nd-order-bound} 
    \end{figure}

    They all satisfy initial conditions $v(0) = v'(0) = 1$. 
    We pick $f(t) =2t^2+8t+7$, $g(t) = t^2+t+3$, and $h(t)=1-2t$, so that the manufactured solution is $v(t) = t^2 + t + 1$ for all three equations.
    Fig. \ref{fig:2nd-order-bound} shows that both $\Bound_{loose}$ (Alg. \ref{alg:single-linear-ode-constant-coeff-loose}) and $\Bound_{tight}$ (Alg. \ref{alg:single-linear-ode-constant-coeff-tight}) strictly bounds the absolute error.
    
\subsection{LINEAR ODE SYSTEM WITH CONSTANT COEFFICIENTS} \label{section:high-dimension}
    \begingroup 
        \setlength\arraycolsep{1pt}
        In this subsection, we train $6$ networks to solve a $6$-dimensional linear system of ODEs with constant coefficients, namely, $\frac{\mathrm{d}}{\mathrm{d}t}\vect{v} + A\vect{v} = \vect{f}$. 
        We pick $A = PJP^{-1}$ where {$J=\begin{pmatrix}J_1\\[-0.75ex]&J_2\\[-0.75ex]&&J_3\end{pmatrix}$} with {$J_1 = \begin{pmatrix} 4&1\\[-0.75ex]&4&1\\[-0.75ex]&&4\\[-0.5ex]\end{pmatrix}$, $J_2 = \begin{pmatrix} 3&1\\[-0.75ex]&3\\[-0.5ex]\end{pmatrix}$ }, $J_3=2$, and $P$ is a random orthogonal matrix.
    \endgroup

    We pick the initial conditions to be {$\vect{v}(0) = P(0\, 0\, 1\, 0\, 1\, 1)^{T}$} and the forcing function to be {$\vect{f}(t) = P(\cos t + 4 \sin t  + \ln(1+t),\, \frac{1}{1+t} + 4 \ln(1+t) + (t+1),\, 4t + 5,\, 2t + 3t^2 + e^t,\, 4 e^t,\, 2 \cos t - \sin t )^T$}, so that the manufactured exact solution is {$\vect{v}(t) = P ( \sin t,\, \ln(t + 1),\, t + 1,\, t^2,\, e^t,\, \cos t)^T$}.

\begin{figure}[!ht]
        \centering
        \includegraphics[width=\linewidth]{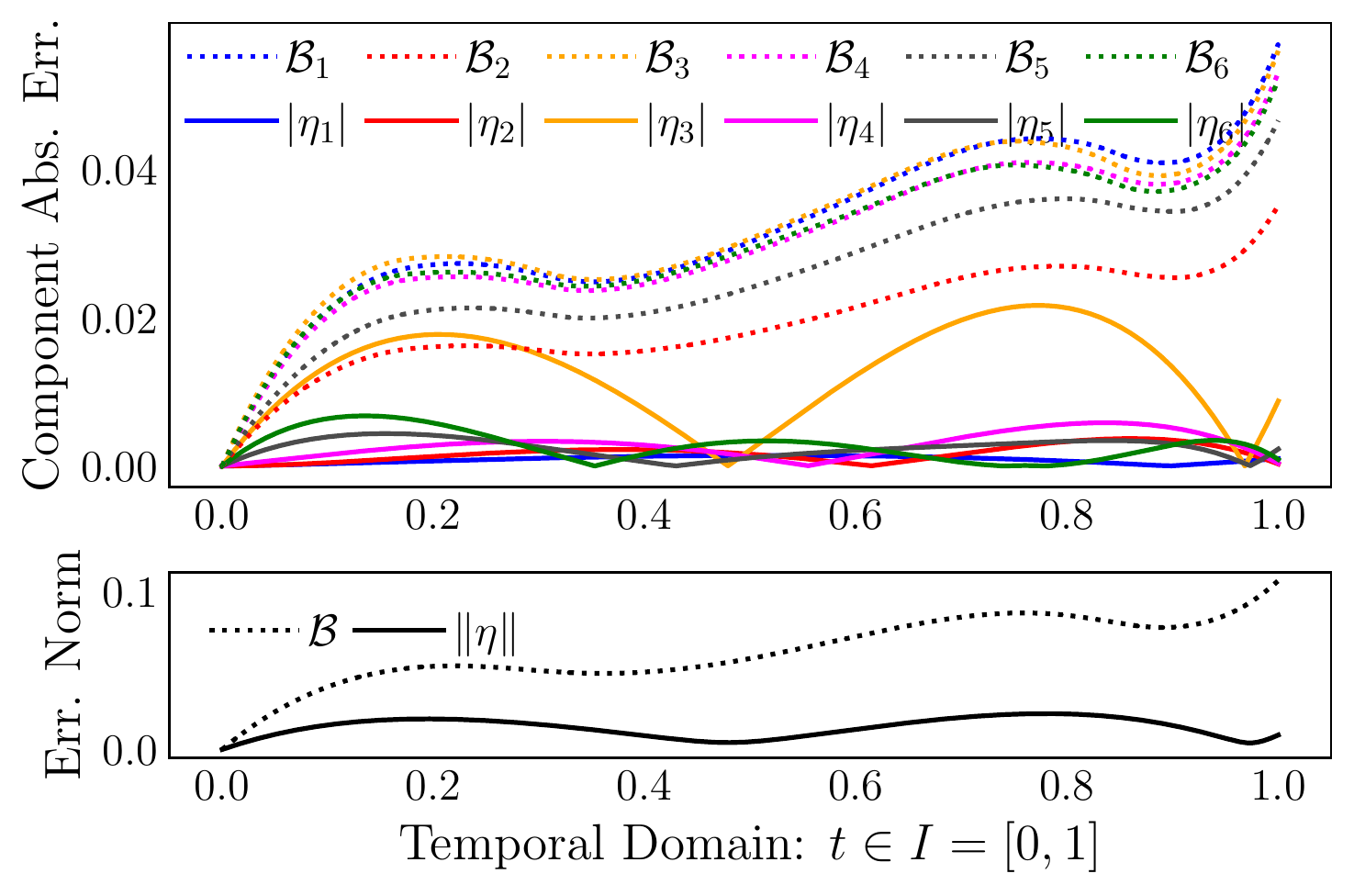}
        \caption{\textit{Componentwise} bound (upper) and \textit{norm} bound (lower) for linear ODE system with constant coefficients}\label{fig:system-bound}
    \end{figure}

    After obtaining the residual information {$\vect{r}(t)=\frac{\mathrm d}{\mathrm d t}\vect{u}(t) + A\vect{u}(t) - \vect{f}(t)$}, we apply Alg. \ref{alg:system-bound} to obtain componentwise bound and norm bound of {$\pmb{\Err} = \vect{u}-\vect{v}$}. 
    It is shown in Fig. \ref{fig:system-bound} that the bounds hold over the domain.

\subsection{NONLINEAR ODE -- DUFFING EQUATION} \label{section:experiment-duffing}
    In this subsection, we consider a Duffing oscillator, which is characterized by the following $2$nd order nonlinear ODE:
        \begin{equation}\label{eq:duffing}
            \frac{\mathrm{d}^2 v}{\mathrm{d}t^2} + 3 \frac{\mathrm{d}v}{\mathrm{d}t} + 2v +\varepsilon v^3 = \cos t ,
        \end{equation}
    under initial conditions $v(0) = 1$ and $v'(0) = 1$, where $\varepsilon$ controls the nonlinearity of the equation. 
    Using Alg. \ref{alg:nonlinear-iterative}, we solve the equation on $I=[0, 2]$ for linspaced $\varepsilon \in (-0.9, 0.9)$ using neural networks and bound the errors. 
    The input $J$ to Alg. \ref{alg:nonlinear-iterative} is chosen to be $6$. 
    Namely, we expand the solution and bound components from degree $0$ to $6$.

\begin{figure}[!ht]
        \centering
        \includegraphics[width=\linewidth]{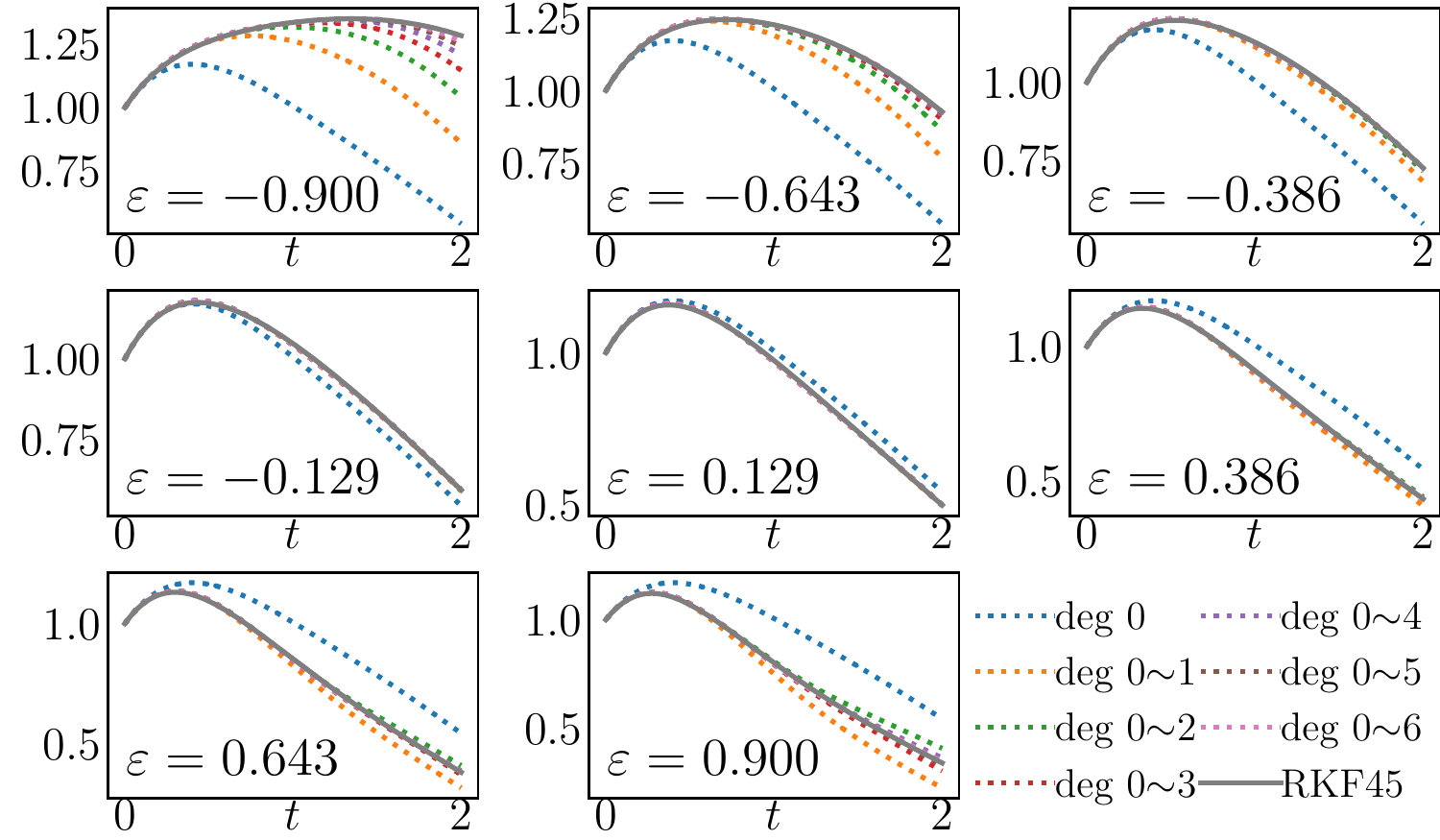}
        \caption{RKF45 and Network Solutions (max-degree 0$\sim$6) to Duffing equation \ref{eq:duffing} for $\varepsilon \in (-0.9, 0.9)$}\label{fig:duffing-solution}
    \end{figure}

\begin{figure}[!ht]
        \centering
        \includegraphics[width=\linewidth]{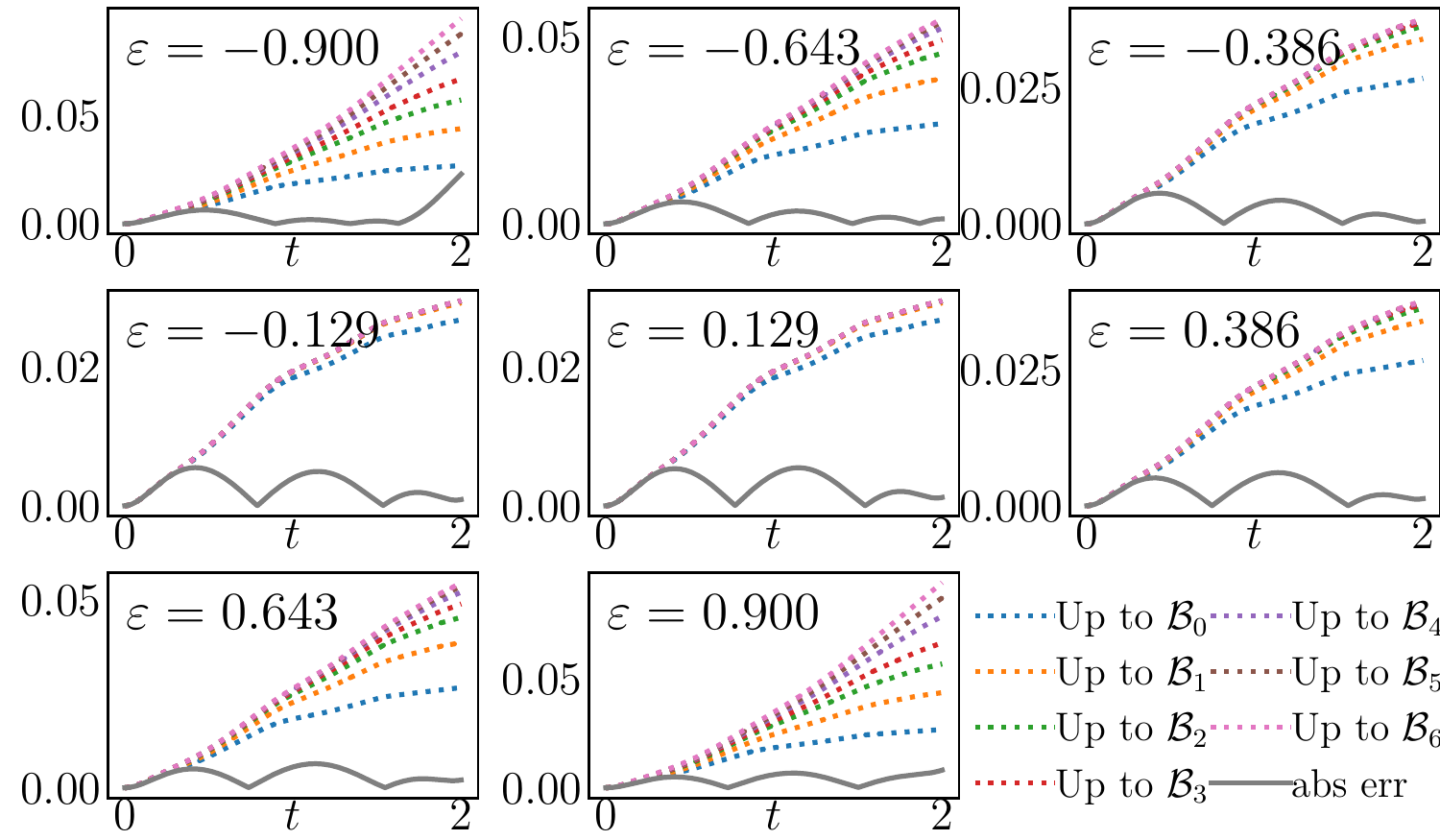}
        \caption{True Error vs. error bound (max-degree 0$\sim$6) of neural network solution to Duffing Equation \ref{eq:duffing} for $\varepsilon \in (-0.9, 0.9)$}\label{fig:duffing-error}
    \end{figure}

    The analytical solution to Eq. \ref{eq:duffing} is complicated. 
    Hence, we use the RKF4(5) method to compute numerical solutions that are close enough to exact solutions for visualization purposes. 
    See Fig. \ref{fig:duffing-solution} for network solutions against RKF4(5) solutions and Fig. \ref{fig:duffing-error} for error bounds against absolute error.

\subsection{LINEAR PDE SYSTEM WITH NONCONSTANT COEFFICIENTS } \label{section:experiment-attractor}
    \begin{figure}[!ht]
        \centering
        \subfloat[\centering $\mathcal{C}: \begin{cases*} x' = -x - y \\ y' = x - y \end{cases*}$]{{\includegraphics[width=0.50\linewidth]{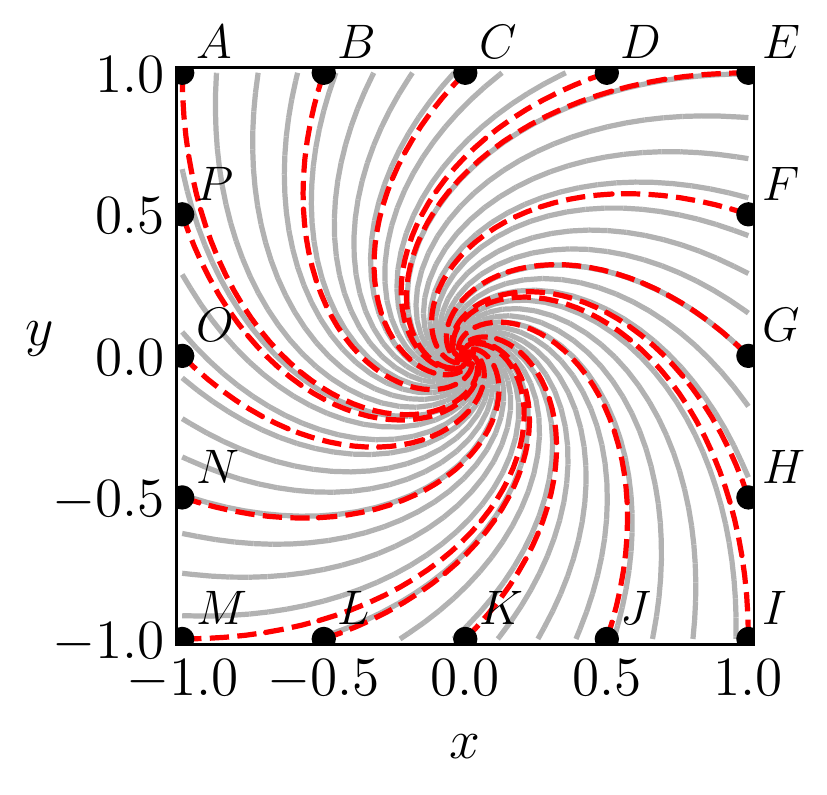}}}
        \subfloat[\centering $\mathcal{C}: \begin{cases*} x' = x^2 + y^2 + 1 \\ y' = x^2 - y^2 + 2 \end{cases*}$]{{\includegraphics[width=0.49\linewidth]{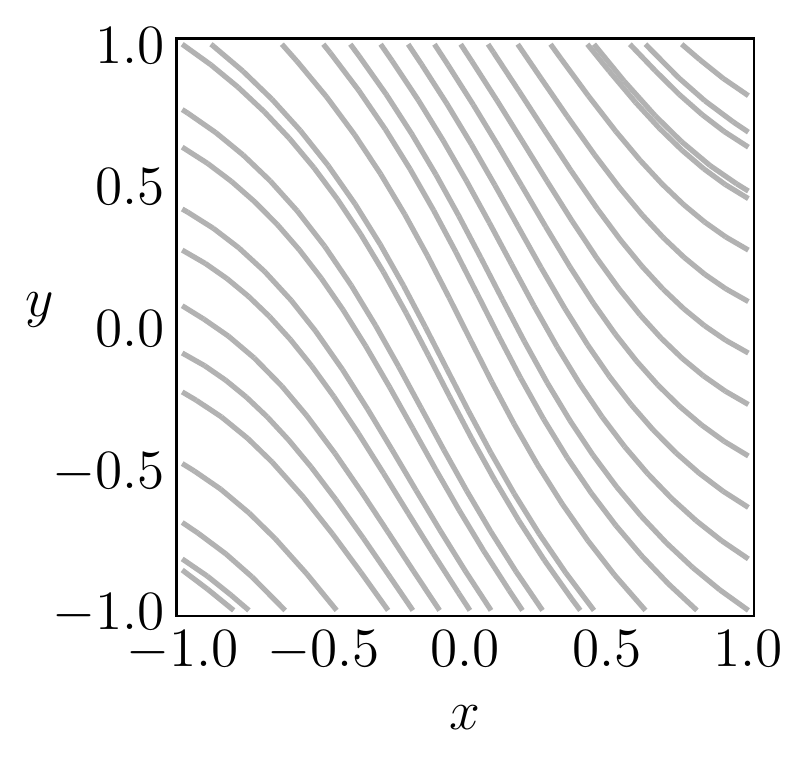}}}
        \caption{
            Characteristics curves of Eq. \ref{eq:attractor} (left) and Eq. \ref{eq:hard-to-solve-characteristics} (right). 
            The red curves, with staring points $A$ to $P$, are selected for visualization of absolute error and error bound in Fig. \ref{fig:pde-error-bound}.
        }\label{fig:characteristics}
    \end{figure}

\subsubsection{PDE Error Bound Evaluation Using Alg. \ref{alg:linear-first-order-pde-general}}
    \begin{figure}[!ht]
        \centering
        \includegraphics[width=\linewidth]{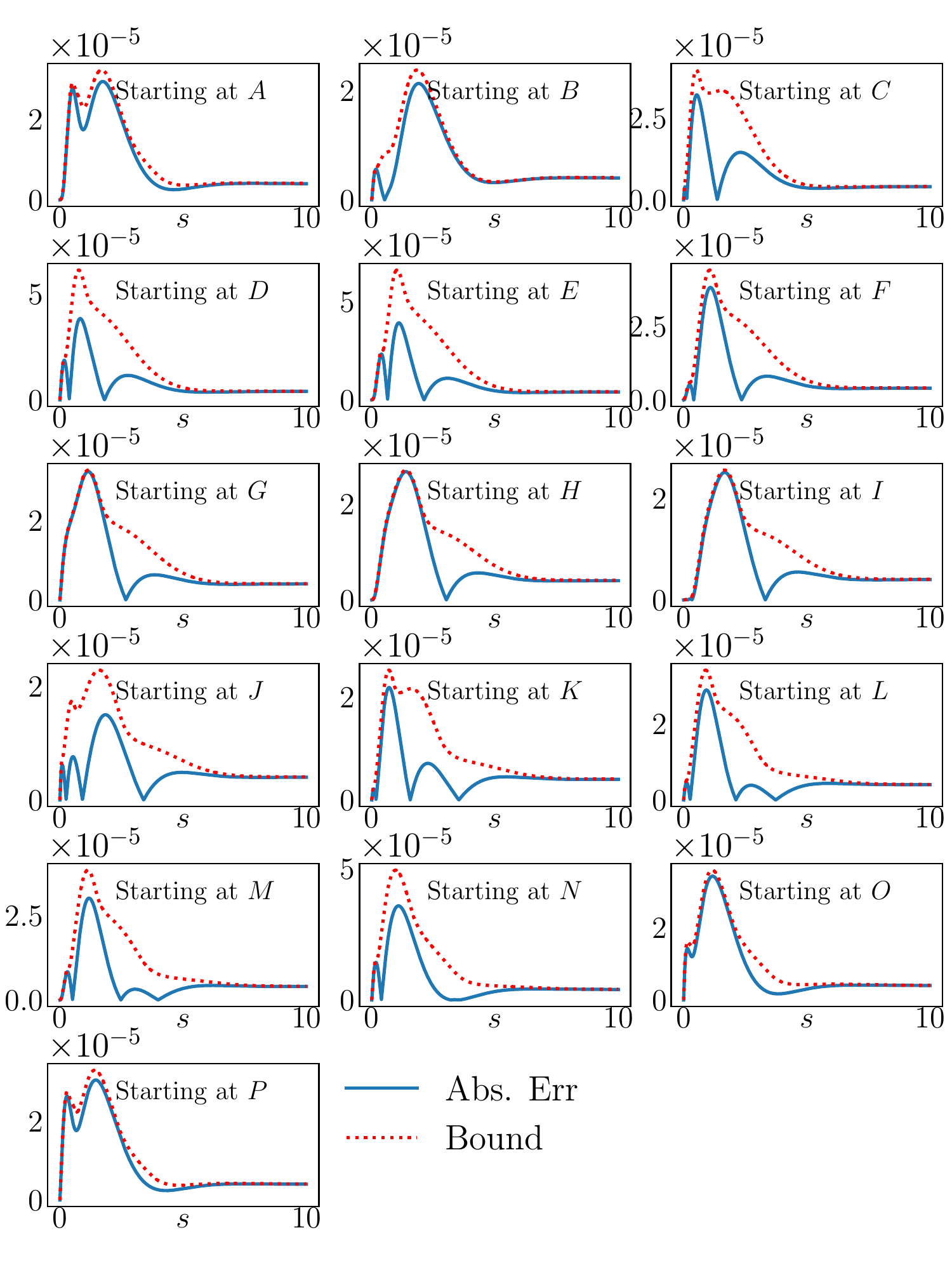}
        \caption{
            Absolute error and error bound on selected characteristic curves. 
            These characteristic curve start at points $A$ through $P$ as shown in Fig. \ref{fig:characteristics}a.
            The blue solid curves are absolute error along the characteristic curves and red dotted curves are corresponding bounds.
        }\label{fig:pde-error-bound}
    \end{figure}

    We try to solve the following first-order linear PDE,
        \begin{equation} \label{eq:attractor}
            (-x -y) \px{v} + (x - y) \py{v} + v = 3x - 2 y
        \end{equation}
    in spatial domain $\Omega=[-1, 1]^2$. 
    The boundary constraints are $v(x, \pm1) =2x\pm 3$ and $v(\pm 1, y) = 3y \pm 2$. 
    The manufactured solution is given by $v(x, y) = 2x + 3y$.
    The characteristic curves are integral curves {$\mathcal{C}: \begin{cases*} x'(s) = -x - y \\[-0.25em] y'(s) = x - y \end{cases*}$}, or {$\mathcal{C}:\begin{cases*} x(s) = R_0 e^{-s} \cos (s+\theta_0)\\[-0.25em] y(s) = R_0 e^{-s} \sin(s + \theta_0) \end{cases*}$}, where $R_0 = \sqrt{x_0^2+y_0^2}$ and $\theta_0 = \mathrm{atan2}(y_0, x_0)$ are constants determined by the starting point $(x_0, y_0) \in \Gamma = \partial \Omega$.
    See Figure \ref{fig:characteristics} for visualization.

    Since the analytical expression of the characteristic curves is known, Alg. \ref{alg:linear-first-order-pde-general} can be applied to evaluate the bound on each curve. 
    We choose $16$ characteristic curves with starting points $A$, $B$, \dots, $P$, equidistantly placed on the boundary (Fig. \ref{fig:characteristics}a). 
    We plot the absolute error and the computed error bound along these characteristic curves in Fig. \ref{fig:pde-error-bound}.
    It can be seen that absolute error lies strictly within the bounds.
\subsubsection{PDE Error Bound Evaluation Using Alg. \ref{alg:linear-first-order-pde-constant}}
    Consider the following PDE 
        \begin{equation}\label{eq:hard-to-solve-characteristics}
            (x^2+y^2+1)\px{v} + (x^2-y^2+2)\py{v} + (3-2x)v = f
        \end{equation}
    over domain $\Omega = [-1, 1]^2$, where $f(x, y) = 6-4x$.
    The boundary constraints are $v(-1, y) = 2$ and $v(x, 1) = 2$, and the manufactured solution is $v(x, y) = 2$.
    
    The characteristic curves {$\mathcal{C}: \begin{cases*} x'(s) = x^2+y^2+1 \\[-0.25em] y'(s) = x^2 - y^2 + 2 \end{cases*}$} are given by a nonlinear ODE, which is hard to solve analytically. 
    (See Fig. \ref{fig:characteristics}b for visualization)
    Therefore, Alg. \ref{alg:linear-first-order-pde-general} cannot be applied to evaluate the error bound. 

\begin{figure}[!ht]
        \centering
        \includegraphics[width=\linewidth]{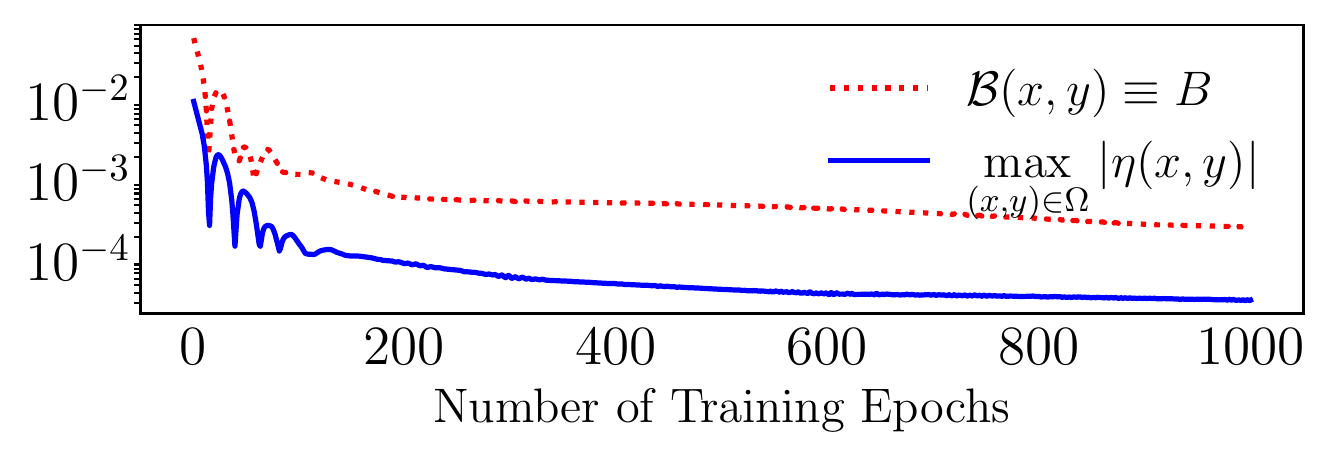}
        \caption{
            Constant bound $B$, computed using Alg. \ref{alg:linear-first-order-pde-constant}, and max absolute error over domain at different epochs of training.
        }\label{fig:pde-constant-bound}
    \end{figure}

    However, the coefficient $(3-2x)$ is nonzero over domain $\Omega$.
    Hence, we can use Alg. \ref{alg:linear-first-order-pde-constant} to compute a constant error bound $|\eta(x, y)| \leq \Bound(x, y) \equiv B$ for all $(x, y) \in \Omega$.
    We visualize the bound and the maximum absolute error $\max_{(x, y)\in\Omega}|\Err|$ after each training epoch in Fig. \ref{fig:pde-constant-bound}.
    As expected, the bound is loose, which is about an order of magnitude larger than the max absolute error.
    Yet, it consistently holds true for every epoch, even during the early stages of training, when the network performs poorly.

\section{CONCLUSION AND FUTURE WORK}
    This paper proposes various error-bounding algorithms for any PINN solution to certain classes of ODEs and PDEs. 
    These algorithms only require the residual information $r(\cdot)$ and the equation structure $\mathcal{D} v = f$ as input.
    There are many real-world applications for which the exact solution $v(\cdot)$ is unknown or hard to compute.
    However, the residual information $r(\cdot)$ is usually, if not always, readily available.
    With our proposed algorithms, PINNs can be trained until the error is gauranteed to fall below a specified tolerance threshold.
    The mathematical relationship between residual and error bound also sheds light on optimizing PINN solutions for future studies.

    The error-bounding algorithms proposed in this paper only apply to certain classes of ODEs and PDEs.
    However, the insights of this paper can be beneficial to future work that extends to more general classes of ODEs and PDEs, especially nonlinear ones.
    We also plan to apply these algorithms stochastic differential equations, where the error bound is a probabilistic tail bound.
    
\bibliography{references}

\end{document}


\onecolumn
\maketitle

\appendix
\section{PROOF OF PROPOSITIONS IN SECTION \ref{section:single-linear-ode-with-constant-coefficients}}
In this part, we first discuss the properties of the operator $\I_{\lambda}$, which is defined in the main paper.
We then use these properties to prove relevant statements regarding Alg. \ref{alg:single-linear-ode-constant-coeff-loose} and Alg. \ref{alg:single-linear-ode-constant-coeff-tight} in Section \ref{section:single-linear-ode-with-constant-coefficients} of the main paper.

\subsection{Properties OF INVERSE OPERATOR $\I_{\lambda} = \L_{\lambda}^{-1}$} \label{appendix:inverse-operator}

    Let $\L_{\lambda}$ ($\lambda \in \mathbb{C}$) be the differential operator $\L_{\lambda}\phi := \dt{\phi} - \lambda \phi$. The inverse of $\L_\lambda \phi = \psi$ is given by $\phi = \I_{\lambda} \psi$ if $\phi(0)=0$, where 
    \begin{equation}
        \I_\lambda \psi (t) := e^{\lambda t}\int_{0}^{t}e^{-\lambda \tau} \psi(\tau)\mathrm{d}\tau.
    \end{equation}

    In addition to $\I_{\lambda} = \L^{-1}_{\lambda}$, there are a few properties of operator $\I_{\lambda}$ that we are interested in
    \begin{enumerate}
        \item \textbf{Linearity:} $\I_{\lambda} (c_1\psi_1 + c_2\psi_2) = c_1\,\I\psi_1 + c_2\,\I\psi_2$ for all functions $\psi_1, \psi_2$ and constants $c_1, c_2 \in \mathbb{C}$
        \item \textbf{Monotonicity:} For $\lambda\in \mathbb{R}$, there is $\big(\forall t\in I, \psi_1(t) \leq \psi_2(t) \big) \Longrightarrow\big(\forall t \in I, \L_{\lambda}\psi_1(t) \leq \L_{\lambda}\psi_2(t)\big)$,
        \item \textbf{Commutativity:} $\I_{\lambda_1} \circ \I_{\lambda_2} = \I_{\lambda_2} \circ \I_{\lambda_1} $ for all $\lambda_1, \lambda_2 \in \mathbb{C}$. This can be shown because $\L_{\lambda_1}\circ\L_{\lambda_2} = \L_{\lambda_2} \circ \L_{\lambda_1}$. Therefore, the inverse operators $\I_{\lambda_2} \circ \I_{\lambda_1} \I_{\lambda_1}\circ\I_{\lambda_2}$ must also be equal.
        \item \textbf{Absolute Inequality:} $|\I_\lambda \psi(t)| \leq \I_{\Re{\lambda}}|\psi(t)|$, which we prove in the next subsection.
    \end{enumerate}

\subsection{Proof of operator inequality $|\I_\lambda \psi| \leq \I_{\Re{\lambda}}|\psi|$}
    \paragraph{Proposition} For any $\lambda \in \mathbb{C}$ and scalar function $\psi: \mathbb{R}^{+} \to \mathbb{C}$, there is 
    \begin{equation}\label{eq:operator-I-inequality}
        |\I_\lambda \psi(t)| \leq \I_{\Re{\lambda}}|\psi(t)|.
    \end{equation}
    \paragraph{Proof} 
    Let $\phi = \I_\lambda \psi$. Since $\L = \I^{-1}$, the problem is equivalent to proving $|\phi| \leq \I_{\Re{\lambda}}|\psi|$, where
    \begin{equation}
        \dt{}\phi-\lambda\phi = \psi.
    \end{equation}
    To see this, we multiply both sides with an integrating factor $e^{-\lambda t}$ and integrate from $0$ to $t$,
    \begin{gather}
        \int_{0}^{t} e^{-\lambda \tau} \left(\frac{\mathrm{d}}{\mathrm{d}\tau}\phi(\tau)-\lambda\phi(\tau)\right)\mathrm{d}\tau = \int_{0}^{t} e^{-\lambda \tau}\psi(\tau) \mathrm{d}\tau\\
        e^{-\lambda t}\phi(t) - \phi(0) = \int_{0}^{t} e^{-\lambda \tau}\psi(\tau) \mathrm{d}\tau
    \end{gather}
    Since $\phi = \I_{\lambda} \psi$, there is $\phi(0) = 0$. Hence we have
    \begin{align}
        \phi(t) &= e^{\lambda t}\int_{0}^{t} e^{-\lambda \tau}\psi(\tau) \mathrm{d}\tau \\
        |\phi(t)| &= \left|e^{\lambda t}\int_{0}^{t} e^{-\lambda \tau}\psi(\tau) \mathrm{d}\tau\right| \\
    \end{align}
    For $\lambda \in \mathbb{C}$, there is $\left|e^{\pm \lambda t}\right| = e^{\pm \Re{\lambda} t}$, where $\Re{\lambda}$ is the real part of $\lambda$.
    Hence,
    \begin{align}
        |\phi(t)| &= e^{\Re{\lambda} t} \left|\int_{0}^{t} e^{-\lambda \tau}\psi(\tau) \mathrm{d}\tau \right| \\
        &\leq e^{\Re{\lambda} t} \int_{0}^{t} \left|e^{-\lambda \tau}\psi(\tau) \right|\mathrm{d}\tau  \\
        &=e^{\Re{\lambda} t} \int_{0}^{t} e^{-\Re{\lambda} \tau}|\psi(\tau)|\mathrm{d}\tau =: \I_{\Re{\lambda}}|\psi(t)|
    \end{align}

\subsection{PROOF OF TIGHT AND LOOSE BOUNDS}
This section proves inequality \ref{eq:single-linear-ode-tight-and-loose} in the main paper, namely,
\begin{equation}\label{eq:tight-bound-to-prove}
    |\Err(t)| \leq \Bound_{tight}(t) :=\left(\I_{\Re{\lambda_1}}\circ\dots\circ\I_{\Re{\lambda_n}}\right)r(t)
\end{equation}
and, if $\Re{\lambda_j} \leq 0$ for all $\lambda_j$, 
\begin{equation}\label{eq:loose-bound-to-prove}
    \Bound_{tight}(t) \leq \Bound_{loose}(t) := \cfrac{1}{Z!} \prod_{\substack{j=1\\ \Re{\lambda_j} \neq 0}}^n \frac{1}{\Re{-\lambda_j}} \max_{\tau\in I}\left|r(\tau)\right|\, t^{Z},
\end{equation}
where $Z$ is the number $\lambda_j$ whose real part is $0$.

\paragraph{Proof} For any linear differential operator $\L=\dnt{n}{} + a_{n-1}\dnt{n-1}{}+\dots+a_0$ whose coefficients $\{a_j\}_{j=0}^{n-1}$ satisfy 
\[
    \lambda^n + a_{n-1}\lambda^{n-1} + \dots +a_0 = \prod_{j=1}^{n} \left(\lambda - \lambda_j\right),
\]
it can be verified that $\L = \L_{\lambda_1} \circ\dots \circ \L_{\lambda_n}$, where $\L_{\lambda_j}\phi := \dt{\phi}-\lambda_j\phi$ as defined in appendix \ref{appendix:inverse-operator}. Then by the proposition in appendix \ref{appendix:inverse-operator}, the inverse operator is given by
\begin{equation}
    \L^{-1} = \left(\L_{\lambda_1} \circ\dots \circ \L_{\lambda_n}\right)^{-1} = \L_{\lambda_n}^{-1} \circ\dots \circ \L_{\lambda_1}^{-1} = \I_{\lambda_n} \circ\cdots\circ\I_{\lambda_1}
\end{equation}
Through repeated application of Inequality \ref{eq:operator-I-inequality}, we can prove Eq. \ref{eq:tight-bound-to-prove}
\begin{align}
    \left|\Err(t)\right| &= \left| \L^{-1} r(t) \right| \\
    &=\left|\left(\I_{\lambda_n} \circ\cdots\circ\I_{\lambda_1}\right) r(t)\right| \\
    &=\left|\I_{\lambda_n}\left(\I_{\lambda_{n-1}} \circ\cdots\circ\I_{\lambda_1}\right) r(t)\right| \\
    &\leq \I_{\Re{\lambda_n}}\left|\left(\I_{\lambda_{n-1}} \circ\cdots\circ\I_{\lambda_1}\right) r(t)\right| \\
    &\leq \left(\I_{\Re{\lambda_n}}\circ \I_{\Re{\lambda_{n-1}}}\right)\left|\left(\I_{\lambda_{n-2}} \circ\cdots\circ\I_{\lambda_1}\right) r(t)\right| \\
    &\leq \dots \nonumber \\
    &\leq \left(\I_{\Re{\lambda_n}}\circ \cdots \circ\I_{\Re{\lambda_{1}}}\right)\left|r(t)\right| =: \Bound_{tight}(t).
\end{align}

In order to prove Eq. \ref{eq:loose-bound-to-prove}, consider the cases of $\Re{\lambda} < 0$ and $\Re{\lambda} = 0$ separately.
\begin{itemize}
    \item If $\Re{\lambda} < 0$, for any constant $c\in \mathbb{R^{+}}$, there is \begin{equation} \I_{\Re{\lambda}} [c] = e^{\Re{\lambda t}} \int_{0}^{t} c e^{-\Re{\lambda}\tau} \mathrm{d}\tau = \frac{c}{-\Re{\lambda}} \left(1 - e^{\Re{\lambda} t}\right) \leq \cfrac{c}{-\Re{\lambda}}\quad \text{for} \quad t\geq0\end{equation}
    \item If $\Re{\lambda} = 0$, for any monomial $c t^{m}$, there is \begin{equation} \I_{\Re{\lambda }}[ct^m] = \I_{0}[ct^m] \int_{0}^{t} c\mathrm{\tau} ^m \mathrm{d}\tau = \frac{c}{m+1}t^{m+1} \quad\text{for}\quad t > 0 \end{equation}
\end{itemize}

Let $\displaystyle R_{\max} := \max_{\tau\in I} |r(t)|$ be the max absolute residual.
Let $Z = |\left\{\lambda_j\ :\Re{\lambda_j} =0,1 \leq j \leq n\right\}|$. 
Assume without loss of generality that $\Re{\lambda_1}, \dots, \Re{\lambda_{n-Z}} < 0$ and that $\Re{\lambda_{n-Z+1}} = \dots = \Re{\lambda_n} = 0$.
By the monotonicity of operator $\I_{\Re{\lambda}}$, there is $\I_{\Re{\lambda}} \phi_1(t) \leq \I_{\Re{\lambda}} \phi_2(t)$ if $\phi_1(t) \leq \phi_2(t)$ for all $t \in I$. Hence, 
\begin{align}
    \Bound_{tight}(t) &= \left(\I_{\Re{\lambda_n}}\circ \cdots \circ\I_{\Re{\lambda_{1}}}\right)\left|r(t)\right| \\
    &\leq \left(\I_{\Re{\lambda_n}}\circ \cdots \circ\I_{\Re{\lambda_{1}}}\right) R_{\max}\\
    &\leq \left(\I_{\Re{\lambda_n}}\circ \cdots \circ\I_{\Re{\lambda_{2}}}\right) \frac{1}{-\Re{\lambda_1}}R_{\max}\\
    &\leq \dots \nonumber \\
    &\leq \left(\I_{\Re{\lambda_n}}\circ \cdots \circ\I_{\Re{\lambda_{n-Z+1}}}\right) \prod_{j=1}^{n-Z}\frac{1}{-\Re{\lambda_j}}R_{\max}\\
    &= \I_0^Z \left[\prod_{\substack{j=1\\ \Re{\lambda_j}\neq 0}}^{n}\frac{1}{-\Re{\lambda_j}}R_{\max}\right]\\
    &= \frac{1}{Z!} \prod_{\substack{j=1\\ \Re{\lambda_j}\neq 0}}^{n}\frac{1}{-\Re{\lambda_j}}R_{\max}\, t^Z =: \Bound_{loose}(t)
\end{align}
which proves Eq. \ref{eq:loose-bound-to-prove}.

\section{PROOF OF PROPOSITIONS IN SECTION \ref{section:system-of-linear-odes-with-constant-coefficients}}
    In this part, we prove relevant statements regarding Alg. \ref{alg:system-bound} in Section \ref{section:single-linear-ode-with-constant-coefficients} of the main paper.

    Consider the problem \ref{eq:linear-system-master} in main paper. 
    The error $\pmb{\Err}$ of the network solution $\vect{u}$ satisfies the equation
    \begin{equation}\label{eq:system-err-equation}
        \dt{} \pmb{\Err} + A\pmb{\Err} = \vect{r}(t) \quad \text{s.t.} \quad \pmb{\Err}(t=0) = \vect{0}
    \end{equation}
    where $\vect{r(t)} = \dt{}\vect{u}(t) + A\vect{u}(t) - \vect{f}(t)$ is the residual vector.

    With the Jordan canonical form \ref{eq:jordan-definition}, we multiply both sides of Eq. \ref{eq:system-err-equation} by $P^{-1}$,
    \begin{gather}
        P^{-1}\dt{}\pmb{\Err} + P^{-1}A \pmb{\Err} = P^{-1}\vect{r}(t) \\
        P^{-1}\dt{}\pmb{\Err} + JP^{-1} \pmb{\Err} = P^{-1}\vect{r}(t) \\
        \dt{}\pmb{\delta} + J \pmb{\delta}  = \vect{q}(t) 
    \end{gather}
    where $\pmb{\delta}(t) := P^{-1}\pmb{\Err}(t)$ and $\vect{q}(t) = P^{-1}\vect{r}(t)$. Recall that $J$ is a Jordan canonical form consisting of $K$ Jordan blocks. Each Jordan block $J_k$ ($1\leq k \leq K$) is an $n_k \times n_k$ square matrix, with eigenvalue $\lambda_k$ on its diagonal and $1$ on its super-diagonal, where $\sum_{k=1}^{K} n_k = n$. Expanding the vector notations, there is 
    \begingroup
        \newcommand{\?}[1]{\multicolumn{1}{c|}{#1}}
        \begin{equation}
            \dt{} 
            \left(\begin{array}{c}
                \delta_{1} \\ \vdots \\ \delta_{n_1} \\ 
                \hline
                \delta_{n_1 + 1} \\ \vdots \\ \delta_{n_1 + n_2} \\ 
                \hline
                \vdots 
            \end{array}\right)
            +
            \left(\begin{array}{c|c|c}
                \begin{matrix} && \\ &J_1& \\ && \end{matrix} & 0 & 0 \\[1.9em]
                \hline
                0 & \begin{matrix} && \\ &J_2& \\ && \end{matrix} & 0 \\[1.9em]
                \hline
                0 & 0 & \ddots 
            \end{array}\right)
            \left(\begin{array}{c}
                \delta_{1} \\ \vdots \\ \delta_{n_1} \\ 
                \hline
                \delta_{n_1 + 1} \\ \vdots \\ \delta_{n_1 + n_2} \\ 
                \hline
                \vdots 
            \end{array}\right)
            =
            \left(\begin{array}{c}
                q_{1}(t)\\\vdots\\ q_{n_1}(t) \\ 
                \hline
                q_{n_1 + 1}(t) \\ \vdots \\ q_{n_1 + n_2}(t) \\ 
                \hline
                \vdots 
            \end{array}\right)
        \end{equation}
    \endgroup

    Let $N_k = n_1+\dots + n_k$. For $k$-th Jordan block indexed by $N_{k-1} < l \leq N_k$, there is
    \begin{equation}
        \dt{}
        \begin{pmatrix}
            \delta_{N_{k-1} + 1} \\[0.8em] \vdots \\[0.8em] \delta_{N_k}
        \end{pmatrix} 
        + 
        \begin{pmatrix}
            \lambda_k & 1 \\
            & \ddots & \ddots \\
            & & \lambda_k & 1\\
            & & & \lambda_k \\
        \end{pmatrix}
        \begin{pmatrix}
            \delta_{N_{k-1} + 1} \\[0.8em] \vdots \\[0.8em] \delta_{N_k}
        \end{pmatrix} 
        =
        \begin{pmatrix}
            q_{N_{k-1} + 1}(t) \\[0.8em] \vdots \\[0.8em] q_{N_k}(t)
        \end{pmatrix},
    \end{equation}
    which can be formulated as the following sequence of scalar equations, also known as \textit{Jordan chains}:
    \begin{alignat}{4}
        &\dt{}\delta_{N_{k-1} + 1} &+&\lambda_k\delta_{N_{k-1} + 1} &=& q_{N_{k-1}+1} &-& \delta_{N_{k-1}+2}, \label{eq:jordan-chain-first}\\
        &\dt{}\delta_{N_{k-1} + 2} &+&\lambda_k\delta_{N_{k-1} + 2} &=& q_{N_{k-1}+2} &-& \delta_{N_{k-1}+3}, \label{eq:jordan-chain-second}\\
        &&&& \vdots \nonumber\\
        &\dt{}\delta_{N_k-1} &+&\lambda_k\delta_{N_k-1} &=& q_{N_k-1} &-& \delta_{N_k}, \label{eq:jordan-chain-second2last}\\
        &\dt{}\delta_{N_k} &+& \lambda_k\delta_{N_k} &=& q_{N_k}. \label{eq:jordan-chain-last}
    \end{alignat}

    The last equation (Eq. \ref{eq:jordan-chain-last}) of the Jordan chain can be used to bound $\delta_{N_k}$ by applying the inequality \ref{eq:operator-I-inequality}, 
    \begin{equation}\label{eq:jordan-chain-bound-last}
        |\delta_{N_k}| = \left|\I_{-\lambda_k}q_{N_k}\right| \leq \I_{-\Re{\lambda_k}} |q_{N_k}|
    \end{equation}
    Applying the inequality \ref{eq:operator-I-inequality} again to Eq. \ref{eq:jordan-chain-second2last}, there is
    \begin{align}
        |\delta_{N_k-1}| &= \left|\I_{-\lambda_k}\left(q_{N_k - 1} + \delta_{N_k}\right)\right| \\
        &\leq \I_{-\Re{\lambda_k}} |q_{N_k - 1} - \delta_{N_k}| \\
        &\leq \I_{-\Re{\lambda_k}} |q_{N_k - 1}| + \I_{-\Re{\lambda_k}} |\delta_{N_k}| \\
        &\leq \I_{-\Re{\lambda_k}} |q_{N_k - 1}| + \I_{-\Re{\lambda_k}}^2 |q_{N_k}|.
    \end{align}
    The first inequality is a direct application of Eq. \ref{eq:operator-I-inequality}. 
    The second inequality is based on linearity of the operator $\I$ and the triangle inequality. 
    The third inequality is obtained by substituting Eq. \ref{eq:jordan-chain-bound-last}.
    Here the superscript in $\I^2$ denotes compositional square $\I^2 = \I\circ\I$.

    By induction, for the $k$-th Jordan block ($N_{k-1} < l \leq N_k$), there is
    \begin{equation}\label{eq:system-scalar-inequality-transformed}
        |\delta_{l}|  \leq \sum_{j=0}^{N_k - l} \I_{-\Re{\lambda_k}} ^ {j+1} |q_{l+j}|
    \end{equation}
    We use this inequality to bound the norm of the error vector, $\left\|\pmb{\Err}\right\|$, as well as absolute value of each component, $\left|\left(\pmb{\Err}\right)_l\right|$. 
\subsection{Componentwise Bound}
    Using matrix notations, Eq. \ref{eq:system-scalar-inequality-transformed} can be rewritten as
    \begin{equation} \label{eq:system-component-inequality-transformed}
        \pmb{\delta}^{\abs} \preceq \pmb{\I}\,\vect{q}^{\abs}
    \end{equation}
    where $\preceq$ denotes componentwise inequality, the superscript $\abs$ denotes componentwise absolute value, and $\pmb{\I}$ is defined as operator matrix $\pmb{\I} := \begin{pmatrix} \vect{I}_1 \\ & \vect{I}_2 \\ && \ddots \end{pmatrix}$ where each $\vect{I}_k = \begin{pmatrix}
        \I_{-\Re{\lambda_k}} & \I_{-\Re{\lambda_k}}^2 & \dots &\I_{-\Re{\lambda_k}}^{n_k} \\[1ex]
        0 & \I_{-\Re{\lambda_k}} & \dots &\I_{-\Re{\lambda_k}}^{n_k-1} \\
        \vdots & \vdots & \ddots & \vdots \\
        0 & 0 & \dots & \I_{-\Re{\lambda_k}}
    \end{pmatrix}$ is an $n_k \times n_k$ upper-triangular block.
    Notice that $(AB)^{\abs} \preceq A^{\abs} B^{\abs}$ for any compatible matrices $A$ and $B$. Recall $\pmb{\delta}(t) = P^{-1}\pmb{\Err}(t)$ and $\pmb{q}(t) = P^{-1} \vect{r}(t)$, there is
    \begin{equation}
        \pmb{\Err}^{\abs} 
        \preceq P^{\abs}\pmb{\delta}^{\abs} 
        \preceq P^{\abs} \pmb{\I} \left[\vect{q}^{\abs} \right]
        \preceq P^{\abs} \pmb{\I} \left[(P^{-1})^{\abs} \vect{r}^{\abs}\right]
    \end{equation}
\subsection{Norm Bound}
    By Eq. \ref{eq:system-component-inequality-transformed}, we have $ \|\pmb{\delta}\| \leq \big\|\pmb{\I} [\|\vect{q}\| \vect{1}]\big\|$, where $\vect{1}$ is $n \times 1$ (constant) column vector whose components are all equal to 1.

    With $\pmb{\Err} = P \pmb{\delta}$ and $\vect{q} = P^{-1}\vect{r}$, there is $\left\|\pmb{\Err}\right\| \leq \left\|P\right\| \|\pmb{\delta}\|$ and $\|\vect{q}\| \leq \left\|P^{-1}\right\| \|\vect{r}\|$, where $\|\cdot\|$ denotes the norm of a vector or the induced norm of a matrix. Consequently,
    \begin{align}
        \left\|\pmb{\Err}(t)\right\| &\leq \left\|P\right\| \|\pmb{\delta}(t)\| \\
        &\leq \|P\|\, \left\|\pmb{\I}\Big[\|\vect{q}(t)\|\vect{1}\Big]\right\|\\
        &\leq \|P\|\, \left\|\pmb{\I}\Big[\|P^{-1}\| \|\vect{r}\|\vect{1}\Big]\right\|\\
        &\leq \|P\|\|P^{-1}\| \left\|\pmb{\I}\Big[\|\vect{r}\|\vect{1}\Big]\right\|\\
        &=\mathrm{cond}(P)\left\|\pmb{\I}\Big[\|\vect{r}(t)\|\vect{1}\Big]\right\| 
    \end{align}
